\documentclass[floats,aps,epsf,amsfonts]{revtex4}
\usepackage[dvips]{graphicx}
\usepackage[colorlinks]{hyperref}
\usepackage{amsmath, amsthm, amssymb}
\usepackage{multirow}
\usepackage{color}

\definecolor{CiteColor}{rgb}{0,0.5,0}
\hypersetup{citecolor=CiteColor}
\definecolor{RefColor}{rgb}{0.55,0,0}
\hypersetup{linkcolor=RefColor}
\definecolor{darkgreen}{rgb}{0.2,0.7,0.2}

\newcommand{\diff}[2]  {\frac{d #1}{d #2}}
\newcommand{\sdiff}[2]  {\frac{d^2 #1}{d #2^2}}
\newcommand{\pdiff}[2]  {\frac{\partial #1}{\partial #2}}
\newcommand{\spdiff}[2] {\frac{\partial^2 #1}{\partial #2^2}}

\renewcommand{\c}{\hskip0.1cm,}
\newcommand{\p}{\hskip0.1cm.}

\renewcommand{\ell}{{\hat{l}}}

\def\etal{\textit{et al.}~}

\begin{document}

\title{Self force on a scalar charge in Kerr spacetime: circular equatorial orbits}
\author{Niels Warburton}
\author{Leor Barack}
\affiliation{School of Mathematics, University of Southampton, Southampton SO17 1BJ, United Kingdom}

\begin{abstract}
We present a calculation of the scalar field self-force (SSF) acting on a scalar-charge particle in a strong-field orbit around a Kerr black hole. Our calculation specializes to circular and equatorial geodesic orbits. The analysis is an implementation of the standard mode-sum regularization scheme: We first calculate the multipole modes of the scalar-field perturbation using numerical integration in the frequency domain, and then apply a certain regularization procedure to each of the modes. The dissipative piece of the SSF is found to be consistent with the flux of energy and angular momentum carried by the scalar waves through the event horizon and out to infinity. The conservative (radial) component of the SSF is calculated here for the first time. When the motion is retrograde this component is found to be repulsive (outward pointing, as in the Schwarzschild case) for any spin parameter $a$ and (Boyer-Lindquist) orbital radius $r_0$. However, for prograde orbits we find that the radial SSF becomes attractive (inward pointing) for $r_0>r_{\rm c}(a)$, where $r_{\rm c}$ is a critical $a$-dependent radius at which the radial SSF vanishes. The dominant conservative effect of the SSF in Schwarzschild spacetime is known to be of 3rd post-Newtonian (PN) order (with a logarithmic running). Our numerical results suggest that the leading-order PN correction due to the black hole's spin arises from spin-orbit coupling at 3PN, which dominates the overall SSF effect at large $r_0$. In PN language, the change-of-sign of the radial SSF is attributed to an interplay between the spin-orbit term ($\propto -ar_0^{-4.5}$) and the ``Schwarzschild'' term ($\propto r_0^{-5}\log r_0$).
\end{abstract}
\date{\today}
\maketitle

\section{Introduction}

The gravitational two-body problem is extremely difficult to tackle in a general-relativistic context, due to the intrinsic nonlinearities of the theory. However, when one of the two components is much more massive than the other the problem simplifies and can sometimes be attacked via black hole perturbation theory. Nature provides us with such extreme mass-ratio systems in the form of compact objects inspiraling into massive black holes in galactic nuclei. Such systems are key targets for the planned space-based gravitational wave detector LISA (Laser Interferometer Space Antenna) \cite{LISA}. Detection of the gravitational waves and accurate extraction of the physical parameters requires precise theoretical templates of the waveforms, which, in turn, necessitate knowledge of the radiative evolution of the system.

The underlying theoretical problem, in its most fundamental form, is that of a pointlike particle orbiting a black hole of a much larger mass. The interaction of the particle with its own gravitational field gives rise to a gravitational self-force (GSF), which is responsible in particular for the radiative inspiral. How to calculate this GSF has been the subject of extensive study over the last decade \cite{Barack-review}. The fundamental formalism for calculations of the GSF in curved spacetime was first laid down by Mino, Sasaki, and Tanaka \cite{Mino-Sasaki-Tanaka} and independently by Quinn and Wald \cite{Quinn-Wald}, with important later supplements by Detweiler and Whiting \cite{Detweiler-Whiting}, Gralla and Wald \cite{Gralla-Wald}, Pound \cite{Pound} and Harte \cite{Harte} (See Poisson for a review \cite{Poisson-review}). The resulting equations of motion are known as the MiSaTaQuWa equations. The analogous self-forced equation of motion for the electromagnetic case was derived by DeWitt and Brehme long ago \cite{DeWitt-Brehme} (with corrections by Hobbs \cite{Hobbs}) and reproduced more recently using other methods in \cite{Quinn-Wald, Gralla-Harte-Wald}. Quinn obtained the equivalent results for the scalar field self-force (SSF) \cite{Quinn}.

The MiSaTaQuWa equations of motion are hard to implement directly and so they were later recast into forms more amenable to practical calculation. One of the standard methods is the mode-sum scheme first introduced in Ref.\ \cite{mode-sum-orig}. Using this method, self force calculations have been performed for a range of problems. These include calculations of the SSF for radial infall \cite{Barack-Burko}, circular \cite{Burko-circular,Diaz-Rivera} and eccentric \cite{Haas} orbits; the electromagnetic self-force for eccentric orbits \cite{Haas-em-eccentric}, and the GSF for radial infall \cite{Barack-Lousto}, circular \cite{Barack-Sago-circular, Detweiler-circular}, and eccentric orbits \cite{Barack-Sago-eccentric}. More recently, researchers have been exploring alternative calculation methods which are based on direct regularization of the self interaction in 2+1 and 3+1 dimensions \cite{Barack-Golbourn-Sago, Lousto-Nakano, Vega}. Common to all calculations presented so far is the fact that they specialize to the simpler (but less astrophysically relevant) case where the central object is a non-rotating, Schwarzschild black hole.

In this paper we open a new front in self force calculations by considering extreme mass-ratio systems where the central black hole is rotating. The motivation for this is clear: Although little is known about the spin distribution of astrophysical massive black holes (but see, e.g., \cite{Lousto-Nakano-Zlochower-Campanelli,Blum-etal}), there is no reason to think that massive holes in nature are non-rotating. Hence, a useful model of a LISA-relevant inspiral must incorporate a Kerr black hole as a central object. Indeed, as this work demonstrates, the spin of the central hole may have a very pronounced effect on the value of the self force and hence on the inspiral dynamics.

Computing the GSF for generic inspiral orbits in Kerr is an extremely challenging task, and this work only represents a first step toward this ultimate goal. The recent advance in calculations of the GSF in Schwarzschild \cite{Barack-Sago-ISCO-shift} was achieved after nearly a decade of development, in which the necessary computational techniques had been devised mainly by using the SSF as a simple test bed. In preparing to tackle the Kerr problem, we once again resort here to the simplicity of the scalar field toy model. Furthermore, as a primer, we specialize to (geodesic) orbits which are both circular and equatorial. This setup already captures much of the complexity of the Kerr problem (and, indeed, offers an opportunity to explore some qualitatively new physics), while providing a more manageable environment for development.

Our calculation represents a first application of the standard mode-sum scheme for orbits in Kerr. As such, it provides a first test of the regularization parameter values derived in Ref.\ \cite{Barack-Ori} (we shall review the notion of regularization parameters in Sec.\ III below). We opt here to work in the frequency domain, with the obvious advantage that we then only need to deal with {\em ordinary} differential equations (ODEs). We decompose the scalar field equation in a basis of spheroidal harmonics (which are frequency-dependent), and solve the resulting ODEs numerically, with suitable boundary conditions. Since the mode-sum scheme requires as input the {\em spherical}-harmonic modes of the scalar field gradient, we then need to re-expand the spheroidal-harmonic solutions into spherical-harmonic components. A major technical hurdle intrinsic to this procedure is that the discontinuity of the spherical-harmonic components across the particle's orbit hampers the convergence of the frequency series there, due to the Gibbs phenomenon. This problem was analyzed in depth in Ref.\ \cite{Barack-Ori-Sago}, and a simple and elegant solution was proposed, which entirely circumvents the problem. With this recent development, the frequency-domain approach becomes an attractive option for SSF studies, in our view. (We remark that the above Gibbs phenomenon issue does not manifest itself in the case of circular orbits considered in our current work.)

In this work we calculate the dissipative and conservative components of the SSF for a variety of orbital radii and black hole spins. Our results for the dissipative component are found to agree well with the numerical results of Gralla \etal \cite{Gralla} (computed from asymptotic fluxes), as well as with the analytic results of Gal'tsov \cite{Galtsov} at large orbital radii. As a further important test of our code we verify that the work done by the dissipative component of the SSF precisely balances the flux of energy in the scalar waves radiated out to infinity and through the event horizon, as extracted from our numerical solutions. For the conservative component our code recovers the results of Diaz-Rivera \etal \cite{Diaz-Rivera} in the Schwarzschild case. This conservative piece is calculated here for the first time for a nonzero Kerr spin parameter, revealing several interesting new features. Our main results for the conservative SSF are displayed in figure \ref{fig:r-a-plane}.

The remainder of this paper is structured as follows. In Sec.\ \ref{section:setup-and-formalism} we
review the relevant features of circular equatorial geodesics of the Kerr geometry, and describe
the setup of our problem. In Sec.\ \ref{section:mode-sum} we discuss the application of the
mode-sum scheme for orbits in Kerr, attempted here for the first time. Section
\ref{section:numerical-implementation} describes our numerical method, and in Sec.\
\ref{section:analysis} we provide various validation tests of our code and present our results.
Lastly in Sec.\ \ref{sec:conclusion} we summarize our results and consider future work. Throughout
this work we use Boyer-Lindquist coordinates $(t,r,\theta, \phi)$, with metric signature $(-+++)$
and geometrized units such that the gravitational constant and the speed of light are equal to
unity.

%%%%%%%%%%%%%%%%%%%%%%%%%%%%%%%%%%%%%%%%%%%%%%%%%%%%%%%%%%%%%%%%%%%%%%%%%%%%%%%%%%%%%%%%%%%%%%%%%%%%%%%%%%%%%%%%%%%%%%%%%%%%%%%%%%%%%%%%%%%%%%%%%%%%%%%%%%%%%%
%%%%%%%%%%%%%%%%%%%%%%%%%%%%%%%%%%%%%%%%%%%%%%%%%%%%%%%%%%%%%%%%%%%%%%%%%%%%%%%%%%%%%%%%%%%%%%%%%%%%%%%%%%%%%%%%%%%%%%%%%%%%%%%%%%%%%%%%%%%%%%%%%%%%%%%%%%%%%%

\section{Setup and review of perturbation formalism}\label{section:setup-and-formalism}

\subsection{Orbit and equation of motion}

Consider a pointlike particle of mass $\mu$ and scalar charge $q$, set in motion about a Kerr black
hole with mass $M$ and spin $aM$. We assume $-M<a<M$, with negative values of $a$ corresponding to
retrograde orbits. We denote the particle's worldline (in Boyer-Lindquist coordinates) by
$x^\mu_p(\tau)$ and its four-velocity by $u^\mu=dx^\mu_p/d\tau$, where $\tau$ is the proper time.
In this work we neglect the GSF, and consider only the SSF, denoted $F^\alpha_{\text{self}}
(\propto q^2)$. Then, the particle's motion is governed by \cite{Quinn}
\begin{eqnarray}
    u^\beta \nabla_\beta (\mu u^\alpha) = F^\alpha_{\text{self}}      \c     \label{eq:sf-geodesic}
\end{eqnarray}
where the covariant derivative is taken (as elsewhere in this work) with respect to the background
Kerr geometry.
In this work we do not wish to consider the back reaction from the SSF on the particle's motion.
Our goal is merely to calculate the SSF that would be felt by a particle fixed on a geodesic orbit.
We envisage that this SSF information could be used to compute the orbital evolution as a second
step, but here we do not attempt carry out the evolution analysis. For simplicity, we specialize to
motion along a geodesic which is both circular [$r_p(\tau)=r_0=$const] and equatorial
[$\theta_p(\tau)\equiv \pi/2$]. Note that, due to the reflective symmetry of the Kerr metric about
the equatorial plane, an initially equatorial orbit (with $\theta_p=\pi/2$ and $d\theta_p/d\tau=0$ at
some initial time) would remain so at all times, even under the influence of the SSF. 

Following from the stationarity and axial symmetry of the background Kerr metric, there exist two Killing vectors,
$\xi_{(t)}^\mu=dx^{\mu}/dt$ and $\xi_{(\phi)}^\mu=dx^{\mu}/d\phi$. 
The Kerr metric also admits a Killing tensor $Q^{\mu\nu}$. To each of these there is associated a conserved
quantity: the specific energy $\mathcal{E}=-\xi_{(t)}^\mu u_\mu =-u_t$, the specific azimuthal angular
momentum $\mathcal{L}= \xi_{(\phi)}^\mu u_\mu = u_\phi$, and the Carter constant $Q= Q^{\mu\nu} u_\mu u_\nu$. Given initial conditions, these three parameters completely specify the orbit of the test particle about the Kerr black hole.

For our circular and equatorial orbits, one readily finds by solving the geodesic equations (taking
$\theta_p=\pi/2$ and $dr_p/d\tau = d^2r_p/d\tau^2 = 0$) \cite{Hughes}
\begin{eqnarray}\label{circular-eng-ang.mom}
\mathcal{E} = \frac{1 - 2v^2 + \tilde{a}v^3}{\sqrt{1 - 3v^2 + 2\tilde{a}v^3}} \c            \quad\quad \mathcal{L} =
r_0v\,\frac{1-2\tilde{a}v^3 + \tilde{a}^2v^4}{\sqrt{1-3v^2+2\tilde{a}v^3}}  \c				\label{eq:energy-and-ang-mom}
\end{eqnarray}
where $v\equiv\sqrt{M/r_0}$ and $\tilde{a}\equiv a/M$. The Carter constant is given explicitly by
\begin{eqnarray}
Q = u_\theta^2 + \cos^2\theta_p\left[a^2(1-\mathcal{E}^2) + \csc^2\theta_p \mathcal{L}^2\right] \c
\end{eqnarray}
and so it vanishes identically in our case.
The angular frequency $\Omega_\phi$ with respect to coordinate time $t$ is given by
\begin{eqnarray}
    \Omega_\phi \equiv \diff{\phi_p}{t} =\frac{u^{\phi}}{u^t}= \frac{g^{\phi\phi}\mathcal{L}-g^{t\phi}\mathcal{E}}{g^{t\phi}\mathcal{L}-g^{tt}\mathcal{E}}= \frac{v^3}{M(1+\tilde a v^3)} \c              \label{eq:orbital-freq}
\end{eqnarray}
where hereafter $g_{\alpha\beta}$ denotes the Kerr background metric, here evaluated at the circular orbit. Notice our convention is that $\mathcal{L}$ and $\Omega_\phi$ are always taken positive, with prograde/retrograde orbits distinguished by the sign of $a$ ($a>0$ for prograde, $a<0$ for retrograde).

Note that in Eq.\ (\ref{eq:sf-geodesic}) we have kept the mass $\mu$ inside the derivative operator. Quinn \cite{Quinn} (see also\ Burko \etal \cite{Burko-Harte-Poisson}) discussed the fact that plausible action principles for the scalar charge in curved spacetime give rise to a dynamically varying mass. In general, the evolution of the mass is governed by the SSF component tangent to $u^{\alpha}$:
\begin{equation}
\frac{d\mu}{d\tau}=-u^{\alpha}F_{\alpha}.
\end{equation} 
In our stationary, circular-orbit setup, however, we must have $d\mu/d\tau=0$. 
Therefore $u^{\alpha}F_{\alpha}=0$ or, more explicitly, 
\begin{equation} \label{F.u=0}
F_t+\Omega_\phi F_\phi=0.
\end{equation}
This trivial relation between $F_t$ and $F_\phi$ means that in our analysis we need only compute one of these components. 

%%%%%%%%%%%%%%%%%%%%%%%%%%%%%%%%%%%%%%%%%%%%%%%%%%%%%%%%%%%%%%%%%%%%%%%%%%%%%%%%%%%%%%%%%%%%%%%%%%%%%%%%%%%%%%%%%%%%%%%%%%%%%%%%%%%%%%%%%%%%%%%%%%%%%%%%%%%%%%
%%%%%%%%%%%%%%%%%%%%%%%%%%%%%%%%%%%%%%%%%%%%%%%%%%%%%%%%%%%%%%%%%%%%%%%%%%%%%%%%%%%%%%%%%%%%%%%%%%%%%%%%%%%%%%%%%%%%%%%%%%%%%%%%%%%%%%%%%%%%%%%%%%%%%%%%%%%%%%

\subsection{Scalar field equation and multipole decomposition}\label{section:setup}

We assume that the particle's field $\Phi$ can be treated as a small perturbation over the fixed Kerr geometry, and that it obeys the minimally coupled Klein-Gordon equation
\begin{equation}\label{eq:fieldeq}
    \nabla_\alpha \nabla^\alpha\Phi  = -4\pi T \c
\end{equation}
sourced by the particle's scalar charge density $T$.  We model this energy-momentum as a $\delta$-function distribution along the particle's worldline, in the form
\begin{eqnarray}\label{T}
T = q\int\delta^4(x^\mu-x^\mu_p(\tau))[-g(x)]^{-1/2}d\tau
  = \frac{q}{r_0^2u^t}\delta(r-r_0)\delta(\phi-\phi_p)\delta(\theta-\pi/2) \c
\end{eqnarray}
where $g=-\rho^4\sin^2\theta$ is the metric determinant, and where in the second equality we have specialized to $r_p=r_0$ and $\theta_p=\pi/2$. The four-velocity component $u^t$ is related to the particle's energy and angular momentum through $u^t= g^{t\phi}\mathcal{L} - g^{tt}\mathcal{E}$.

Carter discovered \cite{Carter} that the scalar wave equation (\ref{eq:fieldeq}) was completely
separable in Kerr geometry, with Brill \etal giving the explicit separation formula \cite{Brill}.
We follow their method and decompose the field into spheroidal harmonics and frequency modes in the
form
\begin{eqnarray}\label{eq:field-decomp}
    \Phi = \int\sum_{\ell=0}^\infty\sum_{m=-\ell}^\ell R_{\ell m\omega}(r)S_{\ell m}(\theta;\sigma^2) e^{im\phi} e^{-i\omega t}\, d\omega   \p
\end{eqnarray}
Here $S_{\ell m}(\theta;\sigma^2)$ are spheroidal Lengendre functions with ($\omega$-dependent) spheroidicity $\sigma^2$ [we reserve the term {\it spheroidal harmonic} for the product $S_{\ell m}(\theta;\sigma^2)e^{im\phi}$]. We label the spheroidal Legendre function by $\ell m$ as we will later introduce {\em spherical} harmonics which we label by $lm$. The spheroidal harmonics are orthonormal with normalization given by
\begin{eqnarray}
    \oint S_{\ell m}(\theta;\sigma^2)e^{im\phi}S_{\ell' m'}(\theta;\sigma^2)e^{-im'\phi} d\Omega = \delta_{\ell\ell'} \delta_{mm'} \c      \label{eq:sh-orthonormal}
\end{eqnarray}
where the integration is over a 2-sphere $t,r=$const with area element $d\Omega=\sin\theta d\theta d\phi$, and $\delta_{n_1n_2}$ is the standard Kronecker delta. 

The source term in Eq.\ (\ref{eq:fieldeq}) is decomposed in a similar manner, writing
\begin{eqnarray}\label{eq:source-decomp}
    \rho^2 T = \int\sum_{\ell=0}^\infty\sum_{m=-\ell}^\ell \tilde{T}_{\ell m\omega}(r)S_{\ell m}(\theta;\sigma^2) e^{im\phi} e^{-i\omega t}\, d\omega  \c
\end{eqnarray}
where the factor $\rho^2\equiv r^2+a^2\cos^2\theta$ is inserted for later convenience.
The periodicity of circular orbits implies that the spectrum of the Fourier transform in Eqs.\ (\ref{eq:field-decomp}) and (\ref{eq:source-decomp}) is given, in our case, by $\omega=n\Omega_\phi\equiv\omega_n$ for integer $n$. Hence for circular equatorial orbits ($r_p=r_0$, $\theta_p=\pi/2$, $\phi_p=\Omega_\phi t$) $ \tilde{T}_{\ell m \omega}$ is given explicitly by
\begin{eqnarray}\label{Tlmw}
        \tilde{T}_{\ell m\omega_n}(r) &= &    \frac{\Omega_\phi}{2\pi}\int_0^{2\pi/\Omega_\phi}{ S_{\ell m}(\theta;\sigma^2)\, \rho^2 T\,  e^{i(n - m)\Omega_\phi t} \, dt}         \nonumber\\
                                    &= &    \frac{q}{u^t} S_{\ell m}(\pi/2;\sigma^2)\delta(r-r_0) \delta^n_{m}  \c
\end{eqnarray}
where in the second line we have substituted for $T$ from
Eq.\ (\ref{T}). Thus, each $m$ mode contains a single $n$-harmonic, and the spectrum is given by
$\omega_n= \omega_m$ with
\begin{equation}
\omega_m \equiv m\Omega_\phi \p
\end{equation}

Substituting the field and source decompositions into the field equation (\ref{eq:fieldeq}) we subsequently find the radial and angular equations to be
\begin{widetext}
\begin{equation}
    \Delta\pdiff{}{r}\left(\Delta\pdiff{R_{\ell m \omega_m}}{r}\right) + \left[a^2m^2-4Mrma\omega_m + (r^2+a^2)^2\omega_m^2 -a^2\omega_m^2\Delta - \lambda_{\ell m}\Delta)\right]R_{\ell m \omega_m} = -4\pi\Delta_0\tilde{T}_{\ell m \omega_m}(r)    \label{eq:radialeq} \c
\end{equation}
\begin{equation}
    \frac{1}{\sin\theta}\pdiff{}{\theta}\left(\sin\theta\pdiff{S_{\ell m}}{\theta}\right) + \left(\lambda_{\ell m} + a^2\omega_m^2\cos^2\theta - \frac{m^2}{\sin^2\theta}\right)S_{\ell m} = 0 \c \label{eq:angulareq}
\end{equation}
\end{widetext}
where $\Delta\equiv r^2-2Mr+a^2$ and $\Delta_0\equiv\Delta(r_0)$. The angular equation (\ref{eq:angulareq}) takes the form of the spheroidal Legendre equation with spheroidicity $\sigma^2=-a^2\omega_m^2$. Its eigenfunctions are the spheroidal Legendre functions $S_{\ell m}(\theta;-a^2\omega_m^2)$ and its eigenvalues are denoted by $\lambda_{\ell m}$. In general there is no closed form for $S_{\ell m}$ or $\lambda_{\ell m}$ but they can be calculated using the spherical harmonic decomposition method described in appendix \ref{apdx:spectral-decomp}.  When $a=0$ the spheroidal harmonics $S_{\ell m}e^{i m \phi}$ coincide with their spherical counterparts $Y_{\hat lm}$ and their eigenvalues reduce to $\lambda_{\ell m} =\ell(\ell+1)$.

As noted by Bardeen \etal \cite{Bardeen} the radial equation (\ref{eq:radialeq}) can be simplified
by transforming to a  new variable,
\begin{eqnarray}
    \psi_{\ell m \omega_m}(r)\equiv rR_{\ell m \omega_m}(r)       \c          \label{eq:R-psi}
\end{eqnarray}
and introducing the tortoise radial coordinate $r_*$ defined through
\begin{equation}\label{eq:rs}
    \diff{r_*}{r} = \frac{r^2}{\Delta} \p
\end{equation}
With the above definition the tortoise coordinate is given explicitly in terms of $r$ as
\begin{eqnarray}
    r_* = r + M \ln(\Delta/M^2) + \frac{(2M^2 - a^2)}{2(M^2-a^2)^{1/2}}\ln\left(\frac{r-r_+}{r-r_-}\right)  \c
\end{eqnarray}
where we have specified the constant of integration and $r_\pm = M \pm \sqrt{M^2-a^2}$ are the
outer and inner roots respectively of the equation $\Delta=0$. We note that there is an alternative
common choice for the tortoise coordinate, namely,
\begin{eqnarray}\label{eq:rs-null}
    \diff{\tilde{r}_*}{r} = \frac{r^2+a^2}{\Delta}  \c
\end{eqnarray}
which is useful in that $v\equiv t+ \tilde{r}_*$ and $u\equiv t- \tilde{r}_*$ are then associated with the ``ingoing'' and ``outgoing'' principal null congruences of the Kerr background \cite{Relativists-Toolkit}. We shall later refer to $\tilde{r}_*$ in discussing boundary conditions, but for our field equation we opt to adopt the coordinate $r_*$, as the $\tilde{r}_*$ coordinate leads to a more complicated radial potential \cite{Brill}.
In terms of $\psi_{\ell m \omega_m}(r)$ and $r_*$, the radial equation (\ref{eq:radialeq}) takes the simpler form
\begin{widetext}
\begin{eqnarray}\label{eq:rsradialeqn}
    \sdiff{\psi_{\ell m \omega_m}}{r_*} + W_{\ell m \omega_m}(r)\psi_{\ell m \omega_m} = -\frac{4\pi q\Delta_0}{r_0^3u^t} S_{\ell m}(\pi/2;-a^2\omega_m^2)\delta(r-r_0) \c
\end{eqnarray}
where we have substituted for the source from Eq.\ (\ref{Tlmw}) and
where $W_{\ell m \omega_m}$ is an effective ($\omega$-dependent) radial potential given by
\begin{eqnarray}
    W_{\ell m \omega_m}(r) = \left[\frac{(r^2+a^2)\omega_m - am}{r^2}\right]^2 - \frac{\Delta}{r^4}\left[\lambda_{\ell m} - 2am\omega_m + a^2\omega_m^2 + \frac{2(Mr-a^2)}{r^2}\right]\p
\end{eqnarray}
\end{widetext}

In the case of circular equatorial orbits, axially-symmetric modes (i.e., ones with $m=0$) have vanishing spheroidicity and $\lambda_{\hat l,m=0}=\hat l(\hat l+1)$. The radial equation (\ref{eq:rsradialeqn}) then admits a simple 
analytic solution. It is given by
\begin{eqnarray}
    \psi_{\ell,m=0} =\left\{ \begin{array}{ll}
            \tilde{\alpha}_{\ell} r Q_\ell(x_0) P_\ell(x)\c \qquad r \le r_0 \c      \\
            \tilde{\alpha}_{\ell}  r P_\ell(x_0) Q_\ell(x)\c \qquad r \ge r_0 \c
            \end{array} \right.
	\label{m=0}
\end{eqnarray}
where
\begin{eqnarray}
    x\equiv \beta (r-M) \quad \text{and}\quad\quad \beta\equiv \sqrt{\frac{M^2+a^2}{M^4-a^4}}   \c
\end{eqnarray}
with $x_0\equiv x(r_0)$ and $P_\ell$ and $Q_\ell$ being the Legendre polynomials of the first and second kind respectively. The coefficient $\alpha_{\ell}$ is derived from the jump condition in the derivative of the field at the location of the particle and is given explicitly by
\begin{eqnarray}
    \tilde{\alpha}_{\ell} = \frac{-4\pi q (u^t \beta \Delta_0)^{-1} S_{\ell 0}(\pi/2;0)}{Q_\ell'(x_0)P_\ell(x_0) - P_\ell'(x_0)Q_\ell(x_0)} \c
\end{eqnarray}
where a prime denotes differentiation with respect to $x$.

\subsection{Boundary conditions}\label{section:bcs}

Equation (\ref{eq:rsradialeqn}) determines the radial field $\psi(r)$ anywhere outside the black hole once
boundary conditions are specified on the horizon ($r_*\to -\infty$) and at spatial infinity
($r_*\to \infty$). The boundary conditions follow from physical considerations: at the event
horizon radiation should be ``ingoing" and at spatial infinity radiation should be ``outgoing" (in
a sense made precise below). As we approach the boundaries the potential $W(r)$ in the radial
equation approaches a constant value and the equation becomes that of a simple harmonic oscillator
with frequencies
\begin{eqnarray}
    W^{1/2}(r_* \rightarrow \infty)   &=& \omega_m      \c                              \\
    W^{1/2}(r_* \rightarrow -\infty) &=& \frac{2Mr_+ \omega_m - am}{r_+^2}\equiv \gamma_m \p
    \label{gamma_m}
\end{eqnarray}
Recalling Eq.\ (\ref{eq:field-decomp}) we observe that, at infinity, the $t,r$-dependence of the
$\ell m \omega$-mode contribution to the full field $\Phi$ will have the asymptotic form
$\Phi_{\ell m \omega}    \sim  \exp[-i\omega_m(t\pm \tilde{r}_*)]/r$, where we have converted from
$r_*$ to $\tilde{r}_*$ by noting that the two coincide (up to an additive constant) at
$r_*\to\infty$. Choosing the sign such that the exponent is expressed in terms of the retarded time coordinate
$u=t-\tilde{r}_*$ ensures that any radiation will be purely outgoing at infinity. Hence the lower sign applies, and the correct asymptotic boundary condition for the radial field is given by
\begin{eqnarray}\label{eqn:field-asymptotics}
    \psi_{\ell m \omega}(r_*\to\infty)    \sim  e^{+i\omega_m r_*} \p     \label{eqn:bcs-infinity}
\end{eqnarray}

At the horizon the situation is slightly more delicate. The asymptotic radial solutions admit the
form $\psi_{\ell m \omega}\sim\exp(\pm i\gamma_m r_*)\sim \exp[\pm i(\omega_m-m\Omega_+) \tilde
r_*]$, where we have expressed $r_*$ in terms of $\tilde r_*$ using the asymptotic relation $r_*\to
[r_{+}/(2M)]\tilde r_*+{\rm const}$ as $r_*\to -\infty$, and defined
\begin{equation}\label{eq:Omega+}
\Omega_+\equiv \frac{a}{2Mr_+} \p
\end{equation}
(The frequency $\Omega_+$ is the angular velocity $\Omega_{\phi}$ of a stationary observer just
outside the event horizon, and might be interpreted as the angular velocity of the black hole
itself \cite{Relativists-Toolkit}.) In evaluating the $\ell m \omega$-mode contribution to the full
field $\Phi$ at the horizon one must now exercise care, and recall that the Boyer-Lindquist
coordinate $\phi$ is {\em singular} at the horizon \cite{Relativists-Toolkit}, and hence the factor
$\exp(im\phi)$ in Eq.\ (\ref{eq:field-decomp}) is singular too. We must instead express the field
in terms of a regular azimuthal coordinate, and, following \cite{Chandra}, we introduce
\begin{equation}
\phi_+\equiv \phi-\Omega_+t.
\end{equation}
In terms of the regular coordinate $\phi_+$ we obtain, as $r_*\to-\infty$, $\Phi_{\ell m \omega}
\sim \exp [im\phi_+ - i(\omega_m-m\Omega_+)(t\mp \tilde{r}_*)]$, where $\mp$ correspond to $\pm$ in
the radial solutions $\psi_{\ell m \omega}\sim\exp(\pm i\gamma_m r_*)$. For this to represent a
purely ingoing radiation the lower sign must be selected, so that $\Phi_{\ell m \omega}$ becomes
asymptotically a function of only $v=t+\tilde r_*$ (as well as the regular angular coordinates
$\phi_+,\theta$). We thus find that the correct boundary condition at the horizon is given by
\begin{eqnarray}\label{eqn:field-asymptotics}
    \psi_{\ell m \omega}(r_*\to -\infty)    \sim  e^{-i\gamma_m {r}_*} \p     \label{eqn:bcs-horizon}
\end{eqnarray}

In passing, we remind that frequency modes with $\omega_m<m\Omega_+$ are {\it superradiant} (see, e.g., Sec. 4.8.2 of \cite{Frolov-Novikov}). Since in our case $\omega_m=m\Omega_{\phi}$, this condition translates to $\Omega_{\phi}<\Omega_+$ [cf.\ Eq.\ (\ref{Edot-}) below] and, using Eqs.\ (\ref{eq:orbital-freq}) and (\ref{eq:Omega+}), also to $r_0>r_0^{\rm sr}(a)$ where $a>0$ and
\begin{equation}
r_0^{\rm sr}(a)\equiv M\left(\frac{r_+^2}{aM}\right)^{2/3}.
\end{equation}
Hence, for prograde circular geodesic orbits with radius greater than $r_0^{\rm sr}(a)$, all
$m$-modes of the scalar field are superradiant. We will demonstrate this behavior numerically in
Sec.\ \ref{subsec:flux} below.

%%%%%%%%%%%%%%%%%%%%%%%%%%%%%%%%%%%%%%%%%%%%%%%%%%%%%%%%%%%%%%%%%%%%%%%%%%%%%%%%%%%%%%%%%%%%%%%%%%%%%%%%%%%%%%%%%%%%%%%%%%%%%%%%%%%%%%%%%%%%%%%%%%%%%%%%%%%%%%
%%%%%%%%%%%%%%%%%%%%%%%%%%%%%%%%%%%%%%%%%%%%%%%%%%%%%%%%%%%%%%%%%%%%%%%%%%%%%%%%%%%%%%%%%%%%%%%%%%%%%%%%%%%%%%%%%%%%%%%%%%%%%%%%%%%%%%%%%%%%%%%%%%%%%%%%%%%%%%

\section{Self force via mode-sum regularization}\label{section:mode-sum}

In the standard mode-sum scheme \cite{mode-sum-orig,Barack-Ori} each vectorial component of the SSF
is constructed from regularized {\it spherical}-harmonic contributions, even in the Kerr case. One
starts by defining the full force as the field
\begin{eqnarray}
    F^{\text{full}}_\alpha(x)\equiv q\nabla_\alpha \Phi(x) = \sum_l F^{\text{(full)}l}_{\alpha}(x)    \c  \label{eq:full-force}
\end{eqnarray}
where $F^{\text{(full)}l}_\alpha$ denotes the total contribution to $\nabla_\alpha\Phi$ from its spherical-harmonic $l$-mode (summed over $m$), and $x$ is shorthand for $x^{\mu}$, an arbitrary field point in the neighbourhood of the particle. Each mode $F^{\text{(full)}l}_{\alpha}$ is finite at the particle's location, although in general the sided limits $r\to r_0^{\pm}$ yield two different values, denoted $F^{\text{(full)}l}_{\alpha\pm}$ respectively. The SSF is then obtained using the mode-by-mode regularization formula
\begin{eqnarray}\label{eq:regularization}
    F_\alpha^{\text{self}} = \sum_{l=0}^\infty \left( F_{\alpha\pm}^{\text{(full)}l} - A_{\alpha\pm} L - B_\alpha \right) \equiv \sum_{l=0}^\infty F^{l\text{(reg)}}_\alpha   \c  \label{eq:Freg}
\end{eqnarray}
where $L\equiv l+1/2$ and the regularized contributions $F^{l\text{(reg)}}_\alpha$ no longer exhibit the $\pm$ ambiguity. The ($l$-independent) {\it regularization parameters} $A_\alpha$ and $B_\alpha$ were first derived for generic orbits about a Schwarzschild black hole \cite{Barack-Mino-Nakano-Ori-Sasaki} and later also for generic orbits in Kerr \cite{Barack-Ori}.
In the circular-equatorial orbit case considered here we have $A_{t\pm}=B_t=0$, and one can show that the mode sum over $F^{l\text{(reg)}}_t$ converges exponentially fast \cite{Barack-review}. For $\alpha=r$ the regularization parameters are generally nonzero and take a rather complicated form; we give these parameters explicitly in appendix \ref{apdx:regularization-params} (specializing to circular equatorial orbits). One usually has $F^{l\text{(reg)}}_r \propto l^{-2}$, so the
mode sum in Eq.\ (\ref{eq:Freg}) converges only as $\sim 1/l$. Recall that one can spare the explicit computation of the $\phi$ component $F_\phi^{\text{self}}$ by using equation (\ref{F.u=0}). Also, from symmetry one obviously has $F_\theta^{\text{self}}=0$ identically.

In Kerr, as we have seen, the scalar field naturally decomposes into spheroidal harmonic modes and hence in order to use the mode-sum scheme in its standard form we must have a preparatory step  where the required spherical-harmonic modes $F_{\alpha\pm}^{\text{(full)}l}$ are to be constructed out of the spheroidal harmonic modes of the scalar field. To achieve this, we first consider the formal expansion of the spheroidal harmonics (with given $\omega$) as a series of spherical harmonics, 
\begin{eqnarray}
    S_{\ell m}(\theta;-a^2\omega_m^2)e^{im\phi} = \sum_{l=0}^\infty b_{lm}^\ell\, Y_{lm}(\theta,\phi)    \c \label{eq:spheroidal-decomp}
\end{eqnarray}
where the coupling coefficients $b_{\ell m}^l=b_{lm}^\ell(a^2\omega_m^2)$ are determined as prescribed in appendix \ref{apdx:spectral-decomp} [this expansion is similar to that applied by Hughes in Ref.\ \cite{Hughes} (with a correction noted by Dolan \cite{Dolan-thesis})]. Note that the spheroidal harmonics and the spherical harmonics have the same $\phi$ dependence (i.e., $e^{im\phi}$) and hence only the $l$ modes couple while the $m$ modes do not. Using Eq.\  (\ref{eq:field-decomp}) in combination with Eqs.\ (\ref{eq:R-psi}) and (\ref{eq:spheroidal-decomp}) we can then express each of the spherical-harmonic $l$-mode contributions in Eq.\ (\ref {eq:full-force}) (for $\alpha=t,r$) in the form
\begin{eqnarray}\label{eq:field-spherical-mode}
    F_\alpha^{\text{(full)}l}(x) = q\nabla_\alpha \sum_{\ell=0}^\infty \sum_{m=-\ell}^\ell b_{l m}^{\ell} \psi_{\ell m}(r) Y_{l m}(\theta,\phi) e^{-i\omega_m t}/r      \p
\end{eqnarray}
The quantities $F_{\alpha\pm}^{\text{(full)}l}$ needed as input for the mode-sum formula (\ref{eq:regularization}) are obtained from the field $F_\alpha^{\text{(full)}l}(x)$ by taking the limits $\theta\to\theta_p$, $\phi\to\phi_p$ and $t\to t_p$, followed by $r\to r_p^{\pm}$.

\begin{figure}
{

        \includegraphics[width=8.5cm]{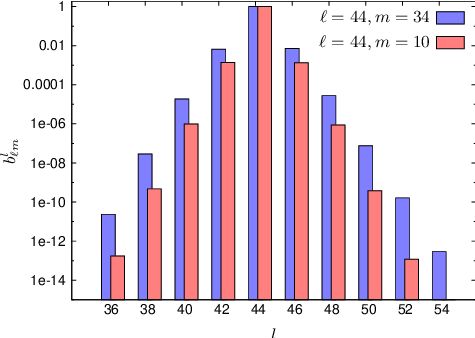}
    }
	{
        \includegraphics[width=8.5cm]{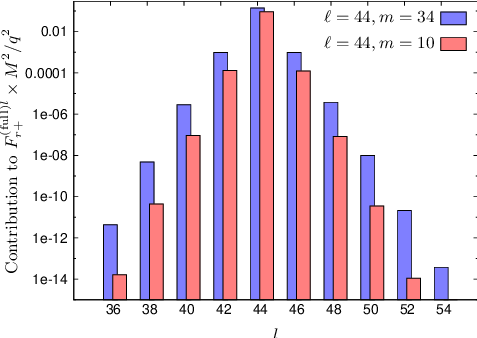}
    }
        \caption[]{Coupling of spheroidal and spherical modes, illustrated here for $a=0.9M$ and $r_0=4M$. Shown are the contributions from a given $\hat{l}$-mode to $b_{lm}^{\hat l}$ (left panel) and $F_{r+}^{\text{(full)}l}$ (right panel), for various spherical harmonic $l$-modes. (Note that $b_{lm}^{\hat l}=0$ identically for odd values of $l-\hat l$.) The width of the $l$ distribution depends mainly on the magnitude of the spheroidicity parameter, $|\sigma^2|=a^2\omega^2=a^2m^2\Omega_{\phi}^2$; the two cases shown, $(\hat{l},m)=(44,34)$ and $(\hat{l},m)=(44,10)$, have spheroidicities $\sigma^2=-11.821$ and $-1.022$, respectively.  The point of this illustration is to note that in practice one only needs to calculate a handful more spheroidal $\ell$ modes than the desired maximum spherical $l$ mode, especially as for smaller $a$ and/or larger $r_0$ the coupling is weaker than demonstrated above.}
\label{fig:field-mode-coupling}
\end{figure}

Note in Eq.\ (\ref{eq:field-spherical-mode}) that whilst formally one must sum over all $\ell$ to construct $F^{\text{(full)}l}_{\alpha}$, in practice this is not necessary as the $\ell$-spectrum (for given $l,m$) is strongly peaked around $\ell=l$; we demonstrate this behavior in figure \ref{fig:field-mode-coupling}. The bandwidth of $\ell$ around $l$ increases slowly with increasing spheroidicity $|\sigma^2|=a^2\omega^2$, yet even at the largest spheroidicity considered in this work ($\sigma^2\sim -126$ for $a=0.998M$, $r_0=2M$), we find that only modes within $l-11\lesssim\ell\lesssim l+11$ carry significant contributions to each of the $l$-modes $F^{\text{(full)}l}_{\alpha}$.

%%%%%%%%%%%%%%%%%%%%%%%%%%%%%%%%%%%%%%%%%%%%%%%%%%%%%%%%%%%%%%%%%%%%%%%%%%%%%%%%%%%%%%%%%%%%%%%%%%%%%%%%%%%%%%%%%%%%%%%%%%%%%%%%%%%%%%%%%%%%%%%%%%%%%%%%%%%%%%
%%%%%%%%%%%%%%%%%%%%%%%%%%%%%%%%%%%%%%%%%%%%%%%%%%%%%%%%%%%%%%%%%%%%%%%%%%%%%%%%%%%%%%%%%%%%%%%%%%%%%%%%%%%%%%%%%%%%%%%%%%%%%%%%%%%%%%%%%%%%%%%%%%%%%%%%%%%%%%

%%%%%%%%%%%%%%%%%%%%%%%%%%%%%%%%%%%%%%%%%%%%%%%%%%%%%%%%%%%%%%%%%%%%%%%%%%%%%%%%%%%%%%%%%%%%%%%%%%%%%%%%%%%%%%%%%%%%%%%%%%%%%%%%%%%%%%%%%%%%%%%%%%%%%%%%%%%%%%
%%%%%%%%%%%%%%%%%%%%%%%%%%%%%%%%%%%%%%%%%%%%%%%%%%%%%%%%%%%%%%%%%%%%%%%%%%%%%%%%%%%%%%%%%%%%%%%%%%%%%%%%%%%%%%%%%%%%%%%%%%%%%%%%%%%%%%%%%%%%%%%%%%%%%%%%%%%%%%

\section{Numerical Implementation}\label{section:numerical-implementation}

For general $\ell$ and $m$ the radial equation (\ref{eq:rsradialeqn}) has no known analytic solutions so it must be solved numerically. To reduce the computation burden one first notes that the individual $\ell m$ modes of the scalar field are invariant under $m \rightarrow -m$ combined with complex conjugation. Consequently when solving the radial equation we need only calculate the modes that have $m \ge 0$ as we can recover the negative $m$ modes by taking the complex conjugate of the corresponding positive $m$ modes.

\subsection{Boundary conditions and junction conditions}\label{section:matching}

The main numerical task is to solve the inhomogeneous radial equation (\ref{eq:rsradialeqn}) with the physical boundary conditions described by (\ref{eqn:bcs-infinity}) and (\ref{eqn:bcs-horizon}). The form of the inner boundary conditions makes it more natural to adopt $r_*$ as the coordinate for the numerical integration. Our numerical domain extends from $r_*=r_{*\text{in}} \ll -M$ out to $r_*=r_{*\text{out}} \gg M$ (how these boundaries are chosen in practice will be discussed below). We assume that the radial field $\psi_{\hat l m}$ admits an asymptotic expansion in $1/r$ at $r\to\infty$ and an asymptotic expansion in $r-r_+$ at $r\to r_+$. Recalling the leading-order behavior of the physical solutions, expressed in Eqs.\ (\ref{eqn:bcs-infinity}) and (\ref{eqn:bcs-horizon}), we thus write
\begin{eqnarray}
    \psi_{\hat l m}(r_{\text{out}}) &=& e^{+ i\omega_m r_{*\text{out}}}\sum^{\bar{k}_{\text{out}}}_{k=0}c^\infty_k r_{\text{out}}^{-k}      \c      \label{eqn:infinity-bc-series}              \\
    \psi_{\hat l m}(r_{\text{in}}) &=& e^{- i\gamma_m r_{*\text{in}}}\sum^{\bar{k}_{\text{in}}}_{k=0}c^{eh}_k(r_{\text{in}}-r_+)^k  \c      \label{eqn:horizon-bc-series}
\end{eqnarray}
where $r_{\text{in}}=r(r_{*\text{in}})$, $r_{\text{out}}=r(r_{*\text{out}})$ and the truncation parameters $\bar{k}_{\text{in,out}}$ are chosen such that the boundary conditions reach a prescribed accuracy (see discussion below). The expansion coefficients are determined by substituting each of the above series into the radial equation. This gives recursion relations for the coefficients $c_{k>0}^{\infty,eh}$ respectively in terms of $c_0^{\infty,eh}$. These relations are rather unwieldy so we relegate their explicit forms to appendix \ref{apx:bcs-recursion-explicit}.

The homogeneous solutions obtained with the above boundary conditions (\ref{eqn:infinity-bc-series}) and (\ref{eqn:horizon-bc-series}) are proportional to the yet-to-be-specified constants $c_0^\infty$ and $c_0^{eh}$ respectively. These constants are determined by imposing suitable matching conditions at the location of the particle. The inhomogeneous solution can be written in the form
\begin{eqnarray}\label{inhom}
    \psi_{\hat l m}(r) = \psi^{-}_{\hat l m}(r)\Theta(r_0-r) +\psi_{\hat l m}^+(r)\Theta(r-r_0) \c
\end{eqnarray}
where $\Theta(x)$ is the Heaviside step function. Substituting this into the radial equation (\ref{eq:rsradialeqn}) and comparing the coefficients of the delta function and its derivative we find
\begin{eqnarray}
    \left.(\psi_{\hat l m}^+  - \psi_{\hat l m}^{-})\right|_{r_0}       &=& 0   \c                          \label{eq:matching-cont}\\
    \left.({\psi_{\hat l m}^+}' - {\psi_{\hat l m}^{-}}')\right|_{r_0}  &=& -\frac{4\pi q r_0}{u^t \Delta_0} S_{\ell m}(\pi/2;-a^2\omega_m^2)\equiv \alpha_{\ell m} \c                  \label{eq:matching-jump}
\end{eqnarray}
where a prime denotes $d/dr$ and, recall, $\Delta_0=\Delta(r_0)$. The first equation implies that the field is continuous at the particle whilst the second describes the nature of the discontinuity in the field's derivative arising from the delta-function source.

In order to determine the correct values of $c_{0}^\infty$ of $c_{0}^{eh}$, for which the conditions (\ref{eq:matching-cont}) and (\ref{eq:matching-jump}) are satisfied, we first numerically solve the radial equation (i) starting from the boundary $r_{\rm out}$ with $c_{0}^\infty=1$ and integrating inward, and (ii) starting from the boundary $r_{\rm in}$ with $c_{0}^{eh}=1$ and integrating outward. We denote the two corresponding homogeneous solutions by $\tilde{\psi}_{\hat l m}^+(r)$ and $\tilde{\psi}_{\hat l m}^{-}(r)$ respectively, so 
\begin{equation}\label{tildepsi}
\psi_{\hat l m}^{+}=c_0^{\infty}\tilde{\psi}_{\hat l m}^{+}\quad \text{and}\quad \psi_{\hat l m}^{-}=c_0^{eh}\tilde{\psi}_{\hat l m}^{-}. 
\end{equation}
Substituting these relations in Eqs.\ (\ref{eq:matching-cont}) and  (\ref{eq:matching-jump}) yields two algebraic equations for $c_0^\infty$ and $c_0^{eh}$, whose solutions read

\begin{eqnarray}
    c_0^{eh} &=& \alpha_{\ell m}\left[\frac{\tilde{\psi}_{\hat l m}^+(r_0)}{\tilde{\psi}_{\hat l m}^{-}(r_0)\tilde{\psi}_{\hat l m}^+{}'(r_0) - \tilde{\psi}_{\hat l m}^+(r_0)\tilde{\psi}_{\hat l m}^{-}{}'(r_0)}\right] \c \label{eq:junc-cond1}\\
    c_0^\infty &=& c^{eh}_0\frac{\tilde{\psi}_{\hat l m}^{-}(r_0)}{\tilde{\psi}_{\hat l m}^+(r_0)} \p           \label{eq:junc-cond2}
\end{eqnarray}
Once the coefficients $c_0^{\infty,eh}$ have been determined, the (unique) physical solution is constructed using Eqs.\ (\ref{inhom}) with (\ref{tildepsi}). 

\subsection{Algorithm}

Following is a summary of the numerical procedure we implement for constructing the SSF. We outline the major steps and give some details about the numerical method and the choice of numerical parameters. 

\begin{itemize}
\item Fix a black hole spin $a$ and orbit radius $r_0$ and calculate the orbital parameters $\mathcal{E}, \mathcal{L}$ and $\Omega_\phi$ [Eqs.\ (\ref{eq:energy-and-ang-mom})and (\ref{eq:orbital-freq})], the spherical harmonic decomposition coefficients $b_{\ell m}^l$ and the spheroidal harmonic eigenvalues $\lambda_{\ell m}$ (the latter two  using the method outlined in appendix \ref{apdx:spectral-decomp}) for all $\hat l$ and $m$ in the range $0 \le \ell \le \ell_{\text{max}} $, $0 \le m \le \ell $. In this work we typically take $\ell_{\text{max}}=55$, which is sufficient for calculating all spherical harmonic contributions $F^{(\rm full)l}_{\alpha\pm}$ up to $l\sim 50$ in most cases; see below. (The estimation of the contribution to the mode-sum from the remaining large-$l$ tail will be discussed in the next subsection.)

\item For each $\hat l$ mode obtain the axially-symmetric mode of the radial variable, $\psi_{\hat l, m=0}$, using the analytic formula (\ref{m=0}). 

\item (For each $m\ne 0$ mode) obtain the boundary conditions for the radial variable using Eqs.\ (\ref{eqn:infinity-bc-series}) and (\ref{eqn:horizon-bc-series}), setting $c_k^{\infty,eh}=1$. Through experimentation we found it practical to set the inner boundary at $r_{*\text{in}}=-60M$. The location of the outer boundary required some adjustment depending on the radius of the particle's orbit. In practice we took $r_{*\text{out}} = 9000M$ for $r_0 < 30M$ and steadily moved it outward for increasing $r_0$ in order to achieve sufficiently fast convergence of the asymptotic series (\ref{eqn:infinity-bc-series}). The largest value for $r_{*\text{out}}$ we used was for $r_0 \geq 100M$ where we had to set $r_{*\text{out}} = 6.0\times10^4M$. We chose $\bar{k}_{\text{in,out}}$ such that the magnitude of the $\bar{k}_{\text{in,out}}+1$ term drops below a certain threshold, which we set to $10^{-14}$.

\item (For each $m\ne 0$ mode) integrate the homogeneous part of the radial equation (\ref{eq:rsradialeqn}) numerically to obtain $\tilde\psi^{\pm}_{\hat lm}(r)$. For this we used the standard Runge--Kutta Prince--Dormand (8,9) method from the GNU Scientific Library (GSL) \cite{GSL}. The GSL Runge-Kutta routine allows one to set a global fractional accuracy target which we took here as $10^{-12}$. To test the integrator we used it to solve for a few $m=0$ modes and compared with the analytic solution (\ref{m=0}). We made further use of the GSL library to calculate many of the special functions (Legendre polynomials, elliptic integrals, Clebsch-Gordan coefficients, etc) that our code requires. 

\item Given the numerical solutions $\tilde\psi^{\pm}_{\hat lm}$ (For each $m\ne 0$ mode), proceed to determine the matching coefficients $c_0^{\infty,eh}$ via Eqs.\ (\ref{eq:junc-cond1}) and (\ref{eq:junc-cond2}), and construct the physical inhomogeneous solutions $\psi_{\hat lm}$ using Eqs.\ (\ref{tildepsi}) and (\ref{inhom}). Record the values of $\psi_{\hat lm}$ and its (one-sided) $r$ and $t$ derivatives at the radius of the particle.

\item Given $\psi_{\hat lm}(r_0)$ and $\nabla_{\alpha\pm}\psi_{\hat lm}(r_0)$ for all spheroidal $\ell m$ modes up to $\ell_{\text{max}}$, use equation (\ref{eq:field-spherical-mode}) to construct the spherical-harmonic $l$ modes of the full force at the location of the particle, $F_{\alpha\pm}^{(\rm full)l}$. 
This procedure allows us to obtain all $l$-modes which do not have significant contributions (through coupling) from the uncalculated modes $\ell>\ell_{\text{max}}$. The highest such $l$ mode, denoted $l_{\text{max}}$, is determined by calculating the contributions from the $\ell_{\text{max}}+1$ spheroidal mode to the various $l$-modes $F_{\alpha\pm}^{(\rm full)l}$, and identifying the highest value of $l$ for which this contribution falls below a given threshold, set here to $10^{-12}$ (fractionally). With $\ell_{\text{max}}=55$ we find $l_{\text{max}}\geq 44$ for all $a$, $r_0$ within the parameter range considered in this work (lower values of $l_{\rm max}$ for larger $|a|$ and smaller $r_0$, with typical values around $l_{\text{max}}\sim 50$); cf.\ figure \ref{fig:field-mode-coupling}. 

\item In the final step, calculate the regularized modes $F_{\alpha}^{l(\rm reg)}$ defined in Eq.\ (\ref{eq:Freg}) using the regularization parameters given in Appendix \ref{apdx:regularization-params}.
Then sum over $l$ modes as in Eq.\ (\ref{eq:Freg}) to obtain the desired SSF. Formally, the mode-sum formula (\ref{eq:Freg}) requires summation over all $l$ modes from $l=0$ to $l=\infty$. In practice, of course, this is neither possible nor necessary. For the $t$ component, the mode sum converges exponentially fast, and we typically find that the contribution from the modes $l\gtrsim 15$ can be safely neglected. For the radial component the situation is a little more subtle, as the mode sum converges only as $\sim 1/l$ in this case---artificially truncating the series at $l\sim 50$ may potentially result in an error of as much as a few tens of percent in the final SSF. It is therefore important to estimate the contribution from the $l>l_{\rm max}$ tail of the mode sum. The method we used for this estimation follows that of Barack and Sago \cite{Barack-Sago-circular}, and for completeness we review it in the next subsection.

\end{itemize}

\subsection{Estimation of the high-$l$ tail contribution}\label{sec:high-l-tail}

We write the total radial component of the SSF as a sum of two pieces, a numerically computed piece, and a large-$l$ tail:
\begin{eqnarray}
    F_r^{\rm self} = F_r^{l\leq l_{\text{max}}} + F_r^{l > l_{\text{max}}} \c
\end{eqnarray}
where, with $F^{l\text{(reg)}}_r$ as defined in equation (\ref{eq:Freg}),
\begin{eqnarray}
    F_r^{l \leq l_{\text{max}}} \equiv \sum^{l_{\text{max}}}_{l=0} F_{r}^{l\text{(reg)}} \qquad \text{and} \qquad F_r^{l>l_{\text{max}}} \equiv \sum^\infty_{l=l_{\text{max}}+1} F_{r}^{l\text{(reg)}} \p
\end{eqnarray}
To evaluate the large-$l$ tail $F^r_{l>l_{\text{max}}}$ we extrapolate the last $\bar n$ numerically calculated $l$-modes using the fitting formula
\begin{eqnarray}
    F_{r}^{l\text{(reg)}} \simeq \sum^N_{n=1}\frac{D^r_{2n}}{L^{2n}} \c
\end{eqnarray}
where, recall, $L=l+1/2$ (how we chose $\bar n$ and $N$ in practice is discussed below). For this fitting we used a standard least-squares algorithm from the GSL. Given the coefficients $D^r_{2n}$, we then estimate the high-$l$ contribution using the formula
\begin{eqnarray}
    F_r^{l>l_{\text{max}}} \simeq \sum^N_{n=1} D^r_{2n} \sum^\infty_{l=l_{\text{max}}+1} L^{-2n} = \sum^N_{n=1} \frac{D^r_{2n}}{(2n-1)!} \Psi_{2n-1}(l_{\text{max}} +3/2) \c
\end{eqnarray}
where $\Psi_n(x)$ is the polygamma function of order $n$ defined as
\begin{eqnarray}
    \Psi_n(x) = \frac{d^{n+1}[\log\Gamma(x)]}{dx^{n+1}} \c
\end{eqnarray}
with $\Gamma(x)$ being the standard gamma function.

Practical use of this estimation method requires some experimentation. For a given $N\in\{3,4,5\}$ we considered a weighted average of the values obtained for $F_r^{\text{self}}$ as we vary $\bar n$ from 20 to 35, where the weighting for each term is given by the square of the inverse of the fractional difference in the value of $F_r^{\text{self}}$ as we increase $\bar n$ by one (this procedure is meant to bias the average in favour of $\bar n$ values for which $F_r^{\text{self}}$ depends only weakly on the number of fitting modes.) We obtain 3 different average values corresponding to $N=3,4,5$, and use the variance of these values to estimate our numerical accuracy (we record as significant figures only those figures that remain fixed as we vary $N$). This error dominates the overall error budget of the SSF, and we hence use it to estimate to overall accuracy of our final SSF results.

It should be noted that the relative contribution from the large $l$ tail is particularly important in the scalar-field case (as compared with the gravitational case). This is because the contribution from the first few $l$ modes turns out to be relatively large and opposite in sign with respect to that of the higher modes. In the Schwarzschild case, the contributions from the $l=0,1$ modes are both negative and (e.g., for $r_0=6M$) conspire to nearly cancel out the combined contributions from $l=3$--$6$. In the Kerr case this cancellation sometimes involves an even greater number of modes (particularly near $a,r_0$ values for which the radial SSF vanishes---see below). This behavior is not observed in the gravitational case \cite{Barack-Sago-circular}---at least not for the Lorenz-gauge GSF in Schwarzschild.

%%%%%%%%%%%%%%%%%%%%%%%%%%%%%%%%%%%%%%%%%%%%%%%%%%%%%%%%%%%%%%%%%%%%%%%%%%%%%%%%%%%%%%%%%%%%%%%%%%%%%%%%%%%%%%%%%%%%%%%%%%%%%%%%%%%%%%%%%%%%%%%%%%%%%%%%%%%%%%
%%%%%%%%%%%%%%%%%%%%%%%%%%%%%%%%%%%%%%%%%%%%%%%%%%%%%%%%%%%%%%%%%%%%%%%%%%%%%%%%%%%%%%%%%%%%%%%%%%%%%%%%%%%%%%%%%%%%%%%%%%%%%%%%%%%%%%%%%%%%%%%%%%%%%%%%%%%%%%

\section{Code validation and results}\label{section:analysis}

\subsection{High-l behavior} \label{sec:large-l}

\begin{figure}
{
        \includegraphics[width=8.0cm]{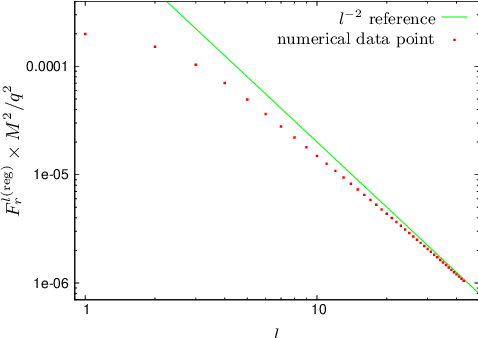}
    }
	{
        \includegraphics[width=8.0cm]{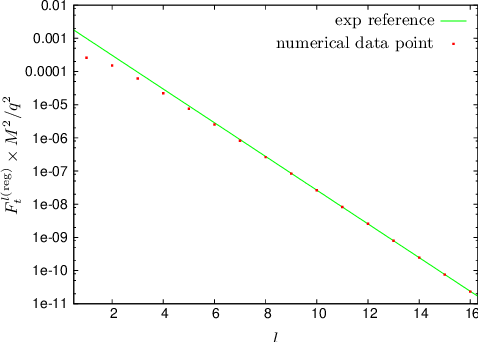}
    }
    \caption[]{{\it Left panel:} the regularized modes $F_r^{l{\rm (reg)}}$ as a function of $l$ for $r_0=5M$ and $a=0.5M$. The solid reference line is $\propto 1/l^2$. The regularized modes demonstrate an asymptotic $\propto 1/l^2$ behavior at large $l$, as expected from theory (note the log-log scale). {\it Right panel:} the regularized modes $F_t^{l{\rm (reg)}}$ as a function of $l$ for $r_0=5M$ and $a=0.8M$. The solid reference line is exponentially decreasing with $l$. The regularized modes of the $t$ component show a clear exponential decay at large $l$, as expected from theory (note the semi-log scale). Similar behavior is observed for other values of $r_0$ and $a$. \label{fig:convergence}  
    }
\end{figure}

According to mode-sum theory \cite{mode-sum-orig}, the regularized modes $F_r^{l{\rm (reg)}}$ in the mode-sum formula (\ref{eq:regularization}) should fall off as $\sim 1/l^2$ for large $l$. This behavior relies sensitively on the delicate cancellation of as many as 3 leading terms in the $1/l$ expansion of the full modes $F_{r\pm}^{{\rm (full)}l}$ (which itself diverges at $\sim l$), and hence provides an excellent test of validity for our numerical results.  Indeed, we have been able to confirm a clear $\sim 1/l^2$  behavior in our numerical data---an example is presented in figure \ref{fig:convergence}. Similarly for the time component, we know from theory that the regularized contributions $F_t^{l{\rm (reg)}}$ decay exponentially with $l$, and again we were able to observe this behavior in our numerical data---see again figure \ref{fig:convergence} for an illustration. The above two tests give us confidence that the high-$\hat l$ spheroidal contributions (whose numerical computation is most demanding) are calculated correctly, and that the spherical-harmonic decomposition procedure is implemented properly. These tests also confirm, for the first time, the validity of the regularization parameters in the Kerr case (for circular equatorial orbits).

\subsection{Energy flux in the scalar waves}\label{subsec:flux}

The above validity check only tests the high-$l$ output of our code. We now discuss a second, more quantitative test, which probes primarily the lower-$l$ portion of the mode sum (and in that sense it is complementary to the first test). From global energy conservation we have that the work done by the dissipative piece (here the $t$ component) of the SSF must be balanced by the flux of energy carried away in scalar-field radiation. We can use our code to compute the flux of energy radiated to infinity and down the black hole, and the result must be consistent with the value of the local dissipative SSF. For the $t$ component the mode-sum converges exponentially fast, and it is for this reason we argued that the energy-balance test is mostly sensitive to the low-$l$ portion of the mode-sum. 

We first briefly review the relevant formalism for computing the radiative flux. The stress-energy tensor of the scalar field is given by
\begin{eqnarray}\label{Tmunu}
    T_{\alpha\beta} = \frac{1}{4\pi}(\Phi_{,\alpha}\Phi_{,\beta} - \frac{1}{2}g_{\alpha\beta}\Phi^{,\mu}\Phi_{,\mu})    \c
\end{eqnarray}
where, as always, $g_{\alpha\beta}$ denotes the Kerr background metric. We wish to consider the flux of scalar-field energy flowing to infinity and down the hole. Let $\Sigma^{+}$ and $\Sigma^{-}$ represent two (timelike) hypersurfaces with $r={\rm const}\gg M$ and $r_*={\rm const}\ll -M$, respectively; and let $d\Sigma^{\pm}$ represent a portion of $\Sigma^{\pm}$ of a small time span $dt$. The amount of scalar-field energy flowing through $\Sigma^{\pm}$ over time $dt$ is expressed by
\begin{equation}\label{flux1}
dE_{\pm}=\mp \oint T^{\alpha}_{\;\;\beta}\xi_{(t)}^{\beta}d\Sigma^{\pm}_{\alpha}
\end{equation}
(see e.g., Sec.\ 4.3.6 of \cite{Relativists-Toolkit}), where $d\Sigma^{\pm}_{\alpha}$ represent outward-pointing surface elements on $d\Sigma^{\pm}$, and the integral is performed over the corresponding 2-spheres of constant $r,t$. The signs are chosen such that the {\em outflow} of energy through $\Sigma^{+}$ is positive, and so is the {\em inflow} of energy through $\Sigma^{-}$ in the Schwarzschild case (recall, however, that $dE_-$ can turn negative in the Kerr case, when superradiance is manifest). In coordinate form we have $\xi_{(t)}^{\beta}=\delta_t^{\beta}$ and $d\Sigma^{\pm}_{\alpha}=(-g^{(3)})^{1/2}\hat r_{\alpha}d\theta d\phi dt=\delta_{\alpha}^{r}\rho^2\sin\theta\, d\theta d\phi dt$, where $g^{(3)}=-\Delta\rho^2\sin^2\theta$ is the determinant of the induced metric on $\Sigma^{\pm}$, and $\hat r_{\alpha}=\delta_{\alpha}^r(g^{rr})^{-1/2}=\delta_{\alpha}^r\Delta^{-1/2}\rho$ is an outward-pointing radial vector of a unit length. The (time-independent) flux of energy through $\Sigma_{\pm}$ is hence given by
\begin{eqnarray}\label{flux2}
    \dot{E}_{\pm} \equiv \diff{E_{\pm}}{t} =  \mp \Delta \oint  T_{tr}\,d\Omega \p
\end{eqnarray}

From Eq.\ (\ref{Tmunu}) we have $T_{tr}=(4\pi)^{-1}\Phi_{,t}\Phi_{,r}$, which, in order to facilitate the angular integration in Eq.\ (\ref{flux2}), we write as $T_{tr}=(4\pi)^{-1}\Phi_{,t}\Phi^*_{,r}$ with an asterisk denoting complex conjugation (this is allowed since $\Phi$ is a real field). We then substitute the spheroidal-harmonic decomposition 
\begin{eqnarray}
\Phi=\frac{1}{r}\sum_{\ell m}\psi_{\ell m}(r) S_{\ell m}(\theta;-a^2\omega_m^2)e^{im\phi}e^{-i\omega_m t} 
\equiv \sum_{\ell m}\Phi_{\ell m} \c
\end{eqnarray}
making the replacement $(\Phi_{\ell m})_{,t}=-i\omega_m\Phi_{\ell m}$. The asymptotic relations 
\begin{eqnarray}
\psi_{\ell m}(r\to\infty)&=&c_0^{\infty}\exp(i\omega_m r) \c \nonumber\\
\psi_{\ell m}(r\to r_+)&=&c_0^{eh}\exp(-i\gamma_m r_*)
\end{eqnarray}
[recall Eqs.\ (\ref{eqn:infinity-bc-series}) and (\ref{eqn:horizon-bc-series})] also allow us to replace 
$(\Phi_{\ell m}^*)_{,r}=-im\Omega_{\phi}\Phi_{\ell m}^*$ for $r\to\infty$, and 
$(\Phi_{\ell m}^*)_{,r}=2iMr_+\Delta^{-1}m(\Omega_{\phi}-\Omega_{+})\Phi_{\ell m}^*$ for $r\to r_+$
[where in the last equality we used Eqs.\ (\ref{eq:rs}), (\ref{gamma_m}) and (\ref{eq:Omega+})].
With these substitutions, the integral in Eq.\ (\ref{flux2}) is readily evaluated using the orthonormality relation (\ref{eq:sh-orthonormal}), giving
\begin{eqnarray} \label{Edot+}
    \dot{E}_{+}   &=& \frac{1}{4\pi}\sum_{\ell m}m^2\Omega_\phi^2 \left|c_0^\infty\right|^2 \c
    \\ 
    \dot{E}_{-}   &=& \frac{M}{2\pi r_+}\sum_{\ell m}m^2\Omega_\phi(\Omega_\phi-\Omega_+) \left|c_0^{eh}\right|^2 \p
    \label{Edot-}
\end{eqnarray}

In Table \ref{table:fluxes} we display numerical values for the total energy flux, $\dot E_{\rm total}\equiv \dot E_+ + \dot E_-$, as computed using our code based on Eqs.\ (\ref{Edot+}) and (\ref{Edot-}). 
For a similar orbital setup, Gralla {\it et al.}\ \cite{Gralla} have previously calculated the total flux of scalar-field {\em angular momentum}, $\dot L_{\rm total}$. In the case of circular, equatorial orbits there applies the simple relation $\dot E_{\rm total} = \Omega_\phi \dot L_{\rm total}$, which allows us a direct comparison with the results obtained in Ref.\ \cite{Gralla}. The data in Table \ref{table:fluxes} shows good agreement between our fluxes and those of Gralla {\it et al.}, with relative differences comparable in magnitude to the estimated relative numerical error in the data of Ref.\ \cite{Gralla}.

\begin{figure}[htb]
    \includegraphics[height=6.5cm]{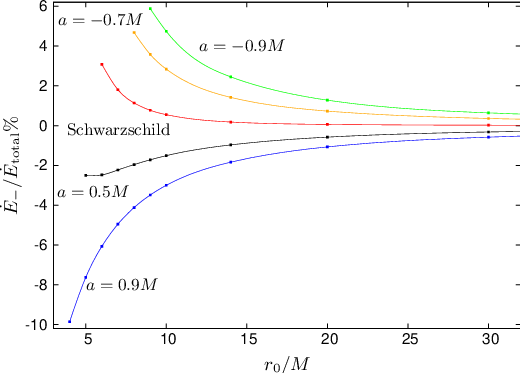}
    \caption{The horizon flux of scalar-field energy, $\dot E_-$, as a percentage of the total flux for different orbital radii $r_0$ and spin parameters $a$. The curves are interpolations based on the numerical data points shown. Superradiance behavior ($\dot E_-<0$) is manifest whenever the horizon's angular velocity $\Omega_+$ is greater than that of the particle.}
    \label{fig:fluxes-ratio}
\end{figure}

Table \ref{table:fluxes} also displays numerical results for the horizon flux, $\dot E_{-}$, expressed as a fraction of  $\dot E_{\rm total}$. Superradiance ($\dot E_{-}<0$) is manifest whenever $\Omega_+>\Omega_\phi$. Horizon absorption does not normally exceed $\sim 10$\% even for strong-field orbits (as also noted by Hughes \cite{Hughes} in the gravitational case), but prograde orbits around a fast rotating hole can display extreme superradiance behavior [nearly 25\%  negative absorption in the example of $(a,r_0)=(0.998M,2M)$]. The graph in figure (\ref{fig:fluxes-ratio}) displays some more horizon absorption data. 

\begin{table}[htb]
    \begin{center}
        \begin{tabular}{|c |c || c c c ||c|}
            \hline
            $a/M$       &   $r_0/M$     &   $q^{-2}\dot{E}_{\text{total}}$    &       $  \dot{E}_-/\dot{E}_{\text{total}}$        &    $1 - \dot{E}_{\text{total}}/\dot{E}_{\text{total}}^{\text{GFW}}$ &    $1 -\dot{E}_{\text{total}}/\dot{\cal E}$   \\
            \hline
            $0.998$   & $2$   	&   $4.3975979e{-}3$ 	&   $-0.2486$ 	&   $7.06e{-}7$  	&	$-4.7e{-}10$    \\
                      & $4$     &   $6.65618888e{-}4$ 	&   $-0.1168$ 	&   $2.12e{-}7$		&	$-1.6e{-}10$    \\
                      & $6$     &   $1.69712483e{-}4$	&   $-0.0692$ 	&   $1.12e{-}6$		&   $-9.2e{-}11$    \\
                      & $8$     &   $6.04494314e{-}5$	&   $-0.0464$ 	&   $-2.12e{-}6$  	&   $-4.6e{-}11$    \\
                      & $10$    &   $2.64845608e{-}5$	&   $-0.0337$ 	&   $6.65e{-}8$  	&   $-3.7e{-}11$    \\
                      & $20$    &   $1.87388789e{-}6$  	&   $-0.0120$ 	&             		&   $-2.8e{-}12$ 	\\
                      & $40$    &   $1.23796212e{-}7$  	&   $-0.0041$ 	&			   		&   $7.7e{-}11$		\\ \hline
            $0.5$     & $6$     &   $2.02918608e{-}4$ 	&   $-0.0248$	&   $-5.19e{-}7$ 	&   $-8.9e{-}11$   	\\
                      & $8$     &   $6.76202950e{-}5$  	&   $-0.0196$ 	&   $-1.76e{-}6$ 	&   $-6.8e{-}11$    \\
                      & $10$    &   $2.86637838e{-}5$	&   $-0.0151$ 	&   $7.33e{-}7$  	&   $-3.3e{-}11$    \\
                      & $20$    &   $1.92605066e{-}6$ 	&   $-0.0058$ 	&               	&   $-1.0e{-}12$    \\
                      & $40$    &   $1.24998716e{-}7$  	&   $-0.0021$ 	&               	&   $-3.5e{-}11$   	\\ \hline
            $0.0$     & $6$     &   $2.55199967e{-}4$ 	&   $0.0308$  	&               	&   $-9.2e{-}11$    \\
                      & $8$     &   $7.72547978e{-}5$ 	&   $0.0114$  	&   $1.98e{-}6$ 	&   $-6.2e{-}11$    \\
                      & $10$    &   $3.13766525e{-}5$ 	&   $0.0054$  	&   $1.28e{-}7$ 	&   $-4.1e{-}11$    \\
                      & $20$    &   $1.98366995e{-}6$  	&   $0.0006$  	&               	&   $-4.6e{-}12$    \\
                      & $40$    &   $1.26226716e{-}7$ 	&   $0.0001$  	&               	&   $4.2e{-}11$  	\\ \hline
            $-0.5$    & $8$     &   $9.02315446e{-}5$	&   $0.0468$  	&   $-5.01e{-}7$  	&   $-4.9e{-}11$   	\\
                      & $10$    &   $3.47579647e{-}5$	&   $0.0284$  	&   $5.20e{-}6$ 	&   $-4.6e{-}11$    \\
                      & $20$    &   $2.04718763e{-}6$ 	&   $0.0073$  	&               	&   $3.4e{-}12$    	\\
                      & $40$    &   $1.27600490e{-}7$	&   $0.0022$  	&               	&   $5.2e{-}11$   	\\ \hline
            $-0.998$  & $9$     &   $6.22560292e{-}5$   &   $0.0644$  	&   $-7.86e{-}7$ 	&   $-5.0e{-}11$    \\
                      & $10$    &   $3.88839360e{-}5$  	&   $0.0519$  	&   $-1.56e{-}6$   	&   $-4.2e{-}11$    \\
                      & $20$    &   $2.11643277e{-}6$   &   $0.0142$  	&               	&   $2.2e{-}11$ 	\\
                      & $40$    &   $1.28992555e{-}7$  	&   $0.0044$  	&               	&   $-6.1e{-}11$ 	\\ \hline
        \end{tabular}
        \caption{Scalar-field energy flux for various values of the spin parameter $a$ and orbital radius $r_0$. The 3rd column displays the total flux of energy radiated to infinity and down the black hole, as extracted from our numerical solutions. The 4th column presents the fraction of the total power absorbed by the black hole, with negative values indicating superradiance. The 5th column compares our fluxes to those obtained by Gralla, Friedman and Wiseman (GFW) \cite{Gralla}, showing a good agreement. (GFW provide results for the radiated angular momentum, which we convert here to radiated energy using the relation $\dot E_{\rm total} = \Omega_\phi \dot L_{\rm total}$; their results are given with 6 significant figures.) In the last column we test our SSF results (for the dissipative component) against the balance relation (\ref{eq:comparison}) as discussed in Sec.\ \ref{subsec:diss}; $\dot{\cal E}(<0)$ is the rate at which the particle's scalar energy is dissipated, as computed from the local SSF using Eq.\ (\ref{local}). 
In this Table (and all subsequent Tables) we use an exponential notation whereby (e.g.) `$e{-}3$' stands for $\times 10^{-3}$. All decimal places presented are significant.        
        }
        \label{table:fluxes}
    \end{center}
\end{table}

\subsection{Dissipative component of the SSF} \label{subsec:diss}

In the case of circular, equatorial orbits the entire information about the dissipative effect of the SSF in contained in the two components $F_t$ and $F_\phi$. Specifically, we obtain from Eq.\ (\ref{eq:sf-geodesic})
\begin{equation}\label{local}
\mu\dot{\cal E}=-(u^t)^{-1}F_t, \quad\quad \mu\dot{\cal L}=(u^t)^{-1}F_\phi,
\end{equation}
where, as elsewhere in this work, an overdot denotes $d/dt$. The relation (\ref{F.u=0}) implies that in practice we need only calculate one of the two components $F_t$ and $F_\phi$---here we choose to calculate the former. Sample numerical data for $F_t$ are presented in table \ref{table:temporal-sf-results}. 

\begin{table}[htb]
    \begin{center}
        \begin{tabular}{|c| c c c c c c c |}
          \hline
          \multicolumn{8}{|c|}{$(M^2/q^2)F_t$} \\
          \hline
            $r_0/M$ & $a=-0.9M$      	& $a=-0.7M$          	& $a=-0.5M$          	& $a=0$           	& $a=0.5M$           		& $a=0.7M$           	& $a=0.9M$       \\
            \hline
            4   &   -              	 	&   -              		&   -              		&   -              		&   -               	&   $1.35921815e{-}3$   &   $1.14204820e{-}3$           \\
            5   &   -              	 	&   -              		&   -              		&   -              		&   $6.07684087e{-}4$   &   $5.35768561e{-}4$   &   $4.79634985e{-}4$           \\
            6   &   -              		&   -              		&   -              		&   $3.60907254e{-}4$  	&   $2.78394798e{-}4$   &   $2.55161013e{-}4$   &   $2.35733853e{-}4$           \\
            7   &   -              	 	&   -              		&   -              		&   $1.76732019e{-}4$	&   $1.46366447e{-}4$   &   $1.37103703e{-}4$   &   $1.29046747e{-}4$           \\
            8   &   -              	 	&   -              		&   $1.15781360e{-}4$   &   $9.77204485e{-}5$	&   $8.44876316e{-}5$   &   $8.02407393e{-}5$   &   $7.64519160e{-}5$           \\
            10  &   $4.60475173e{-}5$   &   $4.38590519e{-}5$   &   $4.18429073e{-}5$   &   $3.75022727e{-}5$   &   $3.40410532e{-}5$   &   $3.28611197e{-}5$   &   $3.17760168e{-}5$           \\
            14  &   $1.03173965e{-}5$   &   $1.00539090e{-}5$   &   $9.80387438e{-}6$   &   $9.23672660e{-}6$   &   $8.74728207e{-}6$   &   $8.57077224e{-}6$   &   $8.40373578e{-}6$           \\
            20  &   $2.28457108e{-}6$   &   $2.25311511e{-}6$   &   $2.22274047e{-}6$   &   $2.15159216e{-}6$   &   $2.08709237e{-}6$   &   $2.06300902e{-}6$   &   $2.03980574e{-}6$           \\
            30  &   $4.30761267e{-}7$   &   $4.27729235e{-}7$   &   $4.24767592e{-}7$   &   $4.17678576e{-}7$   &   $4.11035602e{-}7$   &   $4.08496912e{-}7$   &   $4.06021007e{-}7$           \\
            50  &   $5.43419839e{-}8$   &   $5.41729302e{-}8$   &   $5.40064364e{-}8$   &   $5.36016621e{-}8$   &   $5.32132722e{-}8$   &   $5.30623647e{-}8$   &   $5.29138807e{-}8$           \\
            70  &   $1.40256823e{-}8$   &   $1.39999178e{-}8$   &   $1.39744575e{-}8$   &   $1.39121644e{-}8$   &   $1.38518165e{-}8$   &   $1.38282103e{-}8$   &   $1.38048982e{-}8$           \\
            100 &   $3.35072295e{-}9$   &   $3.34717963e{-}9$   &   $3.34366914e{-}9$   &   $3.33503895e{-}9$   &   $3.32661812e{-}9$   &   $3.32330755e{-}9$   &   $3.32002917e{-}9$           \\
          \hline
        \end{tabular}
\caption{Sample numerical results for the $t$ component of the SSF. Entries left empty correspond to orbits below the inner-most stable circular orbit (ISCO). All figures presented are significant. }
        \label{table:temporal-sf-results}
    \end{center}
\end{table}

In our stationary setting, the rate at which the particle is loosing scalar energy, given by $-\dot{\cal E}$, must equal the rate at which energy flows to infinity and down the black hole, given by $\dot E_{\rm total}$. Using Eq.\ (\ref{local}) we may express this energy balance relation directly in terms of the SSF:
\begin{eqnarray}\label{eq:comparison}
    F_t =   -\mu u^t \dot{\cal E} =  \mu u^t \dot{E}_{\rm total} \p
\end{eqnarray}
As discussed above, this allows us to test our computation of $F_t$ (primarily the low-$l$ portion of the mode-sum) by 
verifying that our numerical results satisfy Eq.\ (\ref{eq:comparison}). As the data presented in right-most column of Table \ref{table:fluxes} demonstrate, we indeed find a very good agreement.

\begin{figure}
    \includegraphics[height=6.5cm]{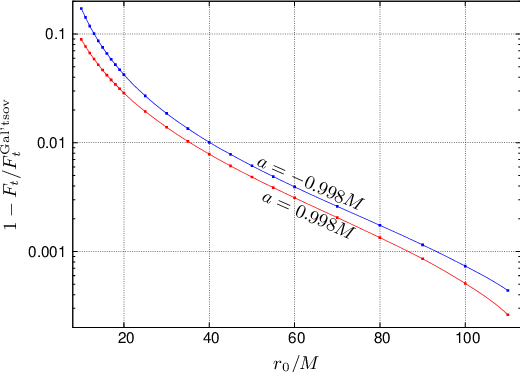}
    \caption{Time component of the SSF: comparison with Gal'tsov's slow-motion formula. Plotted is the relative difference between our ``full'' SSF $F_t$ and Gal'tsov's weak-field/slow motion analytic approximation (\ref{Galtsov}) as a function of orbital radius $r_0$. Solid lines are interpolations of the data points shown. We show results for $a=\pm 0.998M$; similar agreement between $F_t$ and $F_t^{\text{Gal'tsov}}$ at large $r_0$ is manifest for other values of $a$ too. }
    \label{fig:galtsov-comparison}
\end{figure}

It is also interesting to test our results against the weak-field/slow-motion analytic formula derived by Gal'tsov \cite{Galtsov},
\begin{eqnarray}\label{Galtsov}
    F_t^{\text{Gal'tsov}} = \frac{1}{3} q^2 \Omega_\phi \left(r_0^2\Omega^3_\phi + \frac{2M^3 r_+}{r_0^4}(\Omega_\phi - \Omega_+) \right) \c
\end{eqnarray} 
which is valid for $r_0\gg M$.
Here the first term corresponds to the radiation heading out to the infinity and the second to the radiation absorbed by the black hole. In figure \ref{fig:galtsov-comparison} we plot the relative difference between the ``full'' SSF computed here and $F_t^{\text{Gal'tsov}}$ as a function of $r_0$ for a couple of $a$ values (we choose the two extreme cases $a=\pm 0.998M$). Our results seem to obey Gal'tsov's formula for large orbital radii, as expected. 

Lastly, we note that our value of $F_t$ for $(a,r_0)=(0,6M)$ (see Table \ref{table:temporal-sf-results}) coincides through all 9 significant figures with the value computed by Haas and Poisson in Ref.\ \cite{Haas-Poisson-mode-sum}.

\subsection{Conservative component of the SSF}

In our orbital setting, the conservative effect of the SSF is entirely accounted for by its radial component, $F_r$. The computation of this component is more involved, as in this case the mode-sum requires regularization, and (relatedly) the mode-sum series exhibits slow convergence. While results for the dissipative SSF in Kerr (obtained indirectly from the asymptotic fluxes) already exist in the literature, our results for $F_r$ are new. 

\begin{table}[htb]
    \begin{center}
        \begin{tabular}{|c| c c c c c c c |}
          \hline
          \multicolumn{8}{|c|}{$(M^2/q^2)F_r$} \\
          \hline
          $r_0/M$   & $a=-0.9M$         & $a=-0.7M$             & $a=-0.5M$             & $a=0$               & $a=0.5M$               & $a=0.7M$               & $a=0.9M$       \\
          \hline
            $4$ &   -                  &   -                  &   -                  &   -                  &   -                  &   $-5.24194e{-}4$    	&   $-9.5941e{-}4$            \\
            5   &   -                  &   -                  &   -                  &   -                  &   $-4.160235e{-}5$       &   $-2.044174e{-}4$     &   $-3.63448e{-}4$            \\
            6   &   -                  &   -                  &   -                  &   $1.677283e{-}4$    &   $-2.421685e{-}5$       &   $-9.528095e{-}5$     &   $-1.645525e{-}4$          \\
            7   &   -                  &   -                  &   -                  &   $7.850679e{-}5$    &   $-1.467677e{-}5$       &   $-4.980678e{-}5$     &   $-8.410331e{-}5$          \\
            8   &   -                  &   -                  &   $9.642777e{-}5$    &   $4.082502e{-}5$    &   $-9.21907e{-}6$        &   $-2.829488e{-}5$     &   $-4.696081e{-}5$      \\
            10  &   $4.939995e{-}5$    &   $4.100712e{-}5$    &   $3.28942e{-}5$     &   $1.378448e{-}5$    &   $-4.03517e{-}6$        &   $-1.091819e{-}5$     &   $-1.768232e{-}5$        \\
            14  &   $9.968208e{-}6$    &   $8.303689e{-}6$    &   $6.67043e{-}6$     &   $2.720083e{-}6$    &   $-1.07573e{-}6$        &   $-2.561183e{-}6$     &   $-4.02935e{-}6$         \\
            20  &   $1.878548e{-}6$    &   $1.565128e{-}6$    &   $1.2550019e{-}6$   &   $4.93790e{-}7$     &   $-2.50260e{-}7$        &   $-5.43942e{-}7$      &   $-8.35474e{-}7$         \\
            30  &   $2.873310e{-}7$    &   $2.389538e{-}7$    &   $1.90843e{-}7$     &   $7.1719e{-}8$      &   $-4.595209e{-}8$       &   $-9.26682e{-}8$      &   $-1.391883e{-}7$        \\
            50  &   $2.74358e{-}8$     &   $2.272902e{-}8$    &   $1.803392e{-}8$    &   $6.3467e{-}9$      &   $-5.27419e{-}9$        &   $-9.90589e{-}9$      &   $-1.452810e{-}8$        \\
            70  &   $5.87543e{-}9$     &   $4.8525e{-}9$      &   $3.8312e{-}9$      &   $1.2845e{-}9$      &   $-1.25352e{-}9$        &   $-2.26649e{-}9$      &   $-3.27820e{-}9$         \\
            100 &   $1.1508e{-}9$      &   $9.4715e{-}10$     &   $7.4364e{-}10$     &   $2.356e{-}10$      &   $-2.7134e{-}10$        &   $-4.7388e{-}10$      &   $-6.7625e{-}10$         \\
          \hline
        \end{tabular}
        \caption{Sample numerical results for the $r$ component of the SSF. Entries left empty correspond to orbits below the ISCO. All figures presented are significant. The numerical accuracy is lower compared to that of $F_t$ as a result of (i) the regularization procedure involved in obtaining $F_r$, and (ii) the slow decay of the large-$l$ tail in the case of $F_r$ (compared with the exponential decay of the tail for $F_t$). }
        \label{table:radial-sf-results}
    \end{center}
\end{table}

\begin{figure}[htb]
    \includegraphics[height=7.5cm]{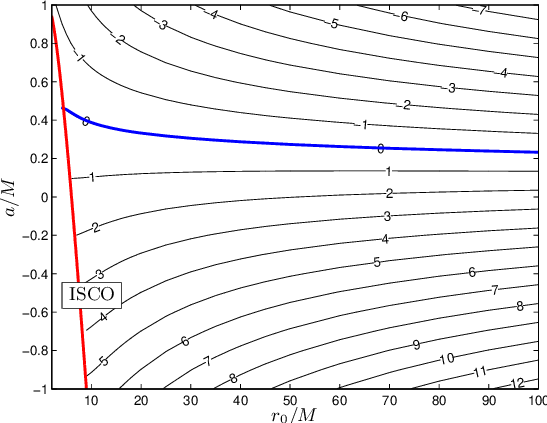}
    \caption{The radial component of the SSF, multiplied by $r_0^5$ for convenience, across the $a,r_0$ parameter space. Contour lines are lines of fixed $r_0^5F_r$, with labels giving the value of $(M/q)^2(r_0/M)^5F_r$. The near-vertical thick line indicates the location of the ISCO, while the near-horizontal thick line marks the curve $r_0=r_c(a)$ along which the radial SSF vanishes. The two lines intersect at $a\simeq 0.461M$; for $a\gtrsim 0.461M$ all stable circular geodesics experience an attractive radial SSF.    
    }
    \label{fig:r-a-plane}
\end{figure}

\begin{figure}[htb]
{
    \includegraphics[width=8.0cm]{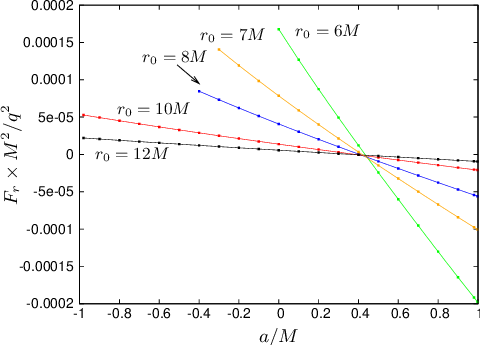}
    }
{
    \includegraphics[width=8.0cm]{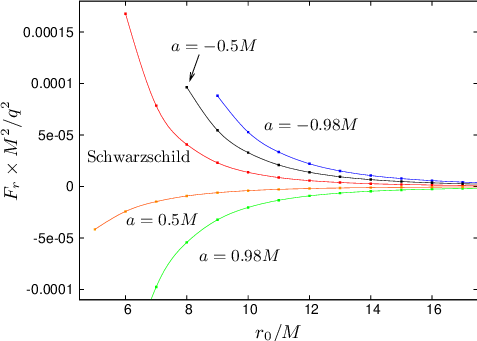}    
    }
    
    \caption{{\it Left panel:} Radial component of the SSF as a function of $a$ for various fixed values of the orbital radius $r_0$.  
   {\it Right panel:} Radial component of the SSF as a function of $r_0$ for various fixed values of the spin parameter $a$. In both panels dots represent numerical data points, and solid lines are interpolations.
    }
    \label{fig:fixed-r}
\end{figure}

Table \ref{table:radial-sf-results} presents $F_r$ data obtained for a range of $a$ and $r_0$ values. Our results for Schwarzschild ($a=0$) agree with those of Diaz-Rivera \etal \cite{Diaz-Rivera} through all significant figures. 
The most striking feature of our results is that---unlike in the Schwarzschild case where the radial SSF is always repulsive (outward pointing)---here we find that for certain prograde orbits $F_r$ becomes {\it attractive} (inward pointing). This behavior is better illustrated in figure \ref{fig:r-a-plane}, where we present a contour plot of $F_r$ across the parameter space of $a,r_0$. This plot is based on the data shown in table \ref{table:radial-sf-results} as well as many other intermediate data points. A few fixed-$r_0$ and fixed-$a$ cross-sections of the contour plot are presented in figure \ref{fig:fixed-r} for clarity. 

We observe the following: (i) For retrograde orbits ($a<0$) the radial SSF is always repulsive, as in the Schwarzschild case. (ii) For prograde orbits ($a>0$) there exists an $a$-dependent radius $r_c$ at which the radial SSF vanishes; it is repulsive for $r_0<r_c$ and attractive for $r_0>r_c$. (iii) The critical radius $r_c$ {\it decreases} monotonically with increasing $a$. (iv) The critical orbit coincides with the ISCO for $a\simeq 0.461M$; hence, all stable circular geodesics experience an attractive radial SSF when $a\gtrsim 0.461M$. It is interesting to note that Burko \cite{Burko-circular} observed a similar change of sign in the radial SSF when studying accelerated (non-geodesic) circular orbits in Schwarzschild geometry.

To gain some intuition about the above behavior of the radial SSF, it is instructive to analyze our results in the context of post-Newtonian (PN) theory. In the Schwarzschild case, a weak-field expression for the radial SSF was worked out to high PN order by Hikida \etal in Ref.\ \cite{Hikida-etal}. Only the leading 3PN and 4PN terms are given explicitly in that work. They read \footnote{Note our definition of the scalar field differs from that of Hikida \etal  \cite{Hikida-etal} by a factor $4\pi$, leading to a similar relative factor in the SSF.}
\begin{eqnarray}\label{Hikida}
    F_r^{(a=0)}(r_0\gg M) = \frac{q^2}{r_0^2}\left[\left(\frac{M}{r_0}\right)^3\left[p_3 + p_3^{\rm log}\ln(r_0/M)\right] + \left(\frac{M}{r_0}\right)^4\left[p_4 + p_4^{\rm log}\ln(r_0/M)\right]\right] \p
\end{eqnarray}
where the coefficient are given by 
\begin{eqnarray}
p_3			 &=& -\frac{4}{3}(\gamma+\ln 2)+\frac{7}{64}\pi^2-\frac{2}{9}=-0.836551\ldots \c	   \nonumber\\
p_3^{\rm log}&=& \frac{2}{3} \c  \nonumber\\
p_4			 &=& -\frac{14}{3}\gamma-\frac{66}{5}\ln 2+\frac{29}{1024}\pi^2+\frac{604}{45} =1.85852\ldots \c \nonumber\\
p_4^{\rm log}&=& \frac{7}{3} \c  
\end{eqnarray}
with $\gamma=0.577215\ldots$ being the Euler number. 
Note the leading 3PN term is dominated by a ``logarithmic running'' term.
Using Eq.\ (\ref{Hikida}) as an ansatz for $a=0$, we performed a two-dimensional fit of a large-$r_0$ subset of our numerical data to a model of the form $F_r = F_r^{(a=0)}+ a{\cal L}\times$ power series in $M/r_0$. We find, at leading order,
\begin{eqnarray} \label{PN}
    F_r(r \gg M) = F_r^{(a=0)} +p_3^{\rm so}\frac{q^2 a\mathcal{L}}{r_0^2} \left(\frac{M}{r_0}\right)^3   \c	\label{eq:Kerr-fit-leading-order}
\end{eqnarray}
with 
\begin{equation}
p_3^{\rm so}\simeq - 1.00091.
\end{equation}
Our numerical accuracy was not sufficient to distinguish between different PN models (including possible logarithmic terms) at higher PN orders, so we do not present here fit results beyond the leading 3PN spin term. This leading term has the interpretation of a spin-orbit coupling (``$\vec{a}\cdot\vec L$''). We are not aware of any explicit analytic calculation of this term in the PN literature. (It might be possible to extract the 3PN spin-orbit term from the formal results of Ref.\ \cite{Damour-Esposito-Far}, which, however, we have not attempted here.) Our numerical fit suggests that the coefficient $p_3^{\rm so}$ of the leading 3PN spin-orbit term is simply $-1$.

\begin{figure}[htb]
    \includegraphics[width=11.0cm]{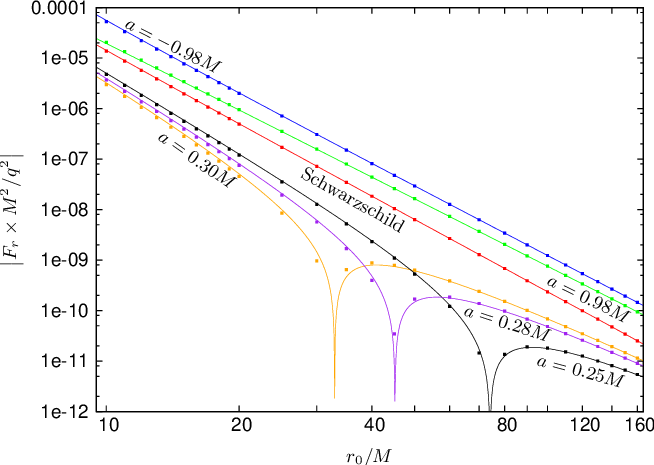}
    \caption{Comparison of numerical data for $F_r$ (dots) with the PN fit model (\ref{PN}) (solid lines). For prograde orbits with $a\lesssim 0.461M$ the radial SSF changes sign at $r_0=r_c(a)$; cf.\ figures \ref{fig:r-a-plane} and \ref{fig:fixed-r}.}
    \label{fig:largeR}
\end{figure}

In figure \ref{fig:largeR} we plot some of our $F_r$ numerical data points against the analytic PN model (\ref{PN}). A good agreement is manifest down to radii as small as $r_0 = 10M$ where the difference between our fitted PN formula (\ref{PN}) and our numerical results is in all cases no more than $8\%$. At $r_0=20M$ this difference is never greater than $3\%$.

We note that $\mathcal{L} \sim r_0^{1/2}$ for large $r_0$ [recall Eq.\ (\ref{circular-eng-ang.mom})], and hence the leading spin term in Eq.\ (\ref{PN}) dominates the overall behavior of $F_r$ at sufficiently large $r_0$, falling off as $\sim r_0^{-4.5}$. At intermediate values of $r_0$, this term, which is negative for $a>0$, competes with the leading ``Schwarzschild'' term, which falls of as $\sim r_0^{-5}\ln r_0$ and is positive. This, we now observe, gives rise to the change-of-sign observed for $F_r$ in our numerical data.

\section{Concluding remarks and future work}\label{sec:conclusion}

In this work we presented a first calculation of the SSF experienced by a particle orbiting a Kerr black hole, specializing to circular and equatorial geodesic orbits. This represented a first application of the mode-sum method in Kerr, and as a by-product we confirmed the analytic values of the regularization parameters $A_{\alpha}$, $B_{\alpha}$ and $C_{\alpha}$, as calculated in \cite{Barack-Ori}, for the above class of orbits. Our numerical calculation relied on a standard frequency-domain decomposition of the scalar field equation in terms of spheroidal harmonics; the spherical-harmonic contributions required within the regularization procedure were obtained by projecting the spheroidal-harmonic contributions onto a basis of spherical harmonics.   

We tested the performance of our code in various ways. The contribution to the SSF from the high-$l$ modes was found to possess the expected behavior, falling off exponentially for the time component and as $\sim l^{-2}$ for the radial component.  We confirmed numerically that the work done by the time component of the SSF precisely balances the energy in scalar waves radiated out to infinity and down through the event horizon. The energy flux calculated from our code also agreed closely with the previous numerical results by Gralla \etal \cite{Gralla} as well as with Galt'sov's analytic formula \cite{Galtsov} in the large-$r_0$ regime. The radial, conservative component of the SSF was calculated here for the first time. Our code produces good agreement with the previous results of Diaz-Rivera \etal \cite{Diaz-Rivera} in the Schwarzschild case. For non-zero spin, we observed a qualitatively new behavior: The SSF 
on prograde orbits with radius larger than a certain $a$-dependent radius $r_c$ turns from repulsive (as in the Schwarzschild case) to attractive. While we have no genuine physical intuition to explain the direction of the radial SSF (not even in the Schwarzschild case), we observed, at a formal level, that the above change-of-sign may be attributed to a competition between a repulsive ``Schwarzschild'' term and an attractive spin-orbit coupling term. 

This observation came from fitting our numerical SSF data to an analytic PN model at large $r_0$. We thus derived a numerical approximation for the leading-order, 3PN spin-correction term. It would be interesting to test our result against an analytic PN computation of the radial SSF, once the PN result becomes available. To further make contact with PN theory it would be necessary to extract higher-order terms in the PN series, and for this it may be necessary  to  improve the accuracy of our code at large orbital radii. The main limiting factor, and by far the dominant source of error in our calculation, is the large contribution to the SSF from the long uncomputed tail of the $l$-mode series. The relative contribution of this tail increases with $r_0$; in our analysis the uncomputed tail contribution for $r_0=100M$ is more than twice that of the computed modes! The problem can be mitigated in future work by pushing our numerical calculation to higher $\ell$, or---better still---by obtaining analytic expressions for some of the higher-order terms in the $1/l$ mode-sum, thereby accelerating the convergence of the mode-sum. (This latter technique was applied successfully by Detweiler \etal in the Schwarzschild case \cite{Detweiler-Messaritaki-Whiting}.) 
 
As mentioned in the introduction, in general a frequency-domain application of the mode-sum method is made difficult by the bad convergence of the frequency mode sum along the particle's orbit (``Gibbs phenomenon''). The problem is unnoticed for circular orbits, since in this case the scalar field is a smooth function of time along the orbit. However, the issue will need to be addressed in contemplating the extension of our code to more generic orbits. The recently introduced method of ``extended homogeneous solutions'' \cite{Barack-Ori-Sago} proposes a simple method to overcome the above difficulty and we envisage incorporating this method in a future extended version of our code. We have already started work to generalize the code to eccentric orbits (which, as a first step, we keep equatorial).

Extension to the gravitational problem is more challenging. The main obstacle is the lack of a formal framework for analyzing Lorenz-gauge metric perturbations in the frequency-domain in Kerr. A potential avenue of approach would be to work with coupled tensorial spherical-harmonics, although this may pose a significant technical challenge. Another possibility would be to develop a suitable tensorial spheroidal-harmonic basis for decomposition in Kerr, akin to the tensorial spherical harmonics that can be used in the Schwarzschild case.

\section*{ACKNOWLEDGEMENTS}

We are grateful to Thibault Damour for a crucial advice relating to our PN fit model. We would also like to thank Sarp Akcay, Steven Detweiler, Sam Dolan, Sam Gralla and Bernard Whiting for helpful comments. NW's work was supported by an STFC grant. LB acknowledges additional support from STFC through grant number PP/E001025/1.

%%%%%%%%%%%%%%%%%%%%%%%%%%%%%%%%%%%%%%%%%%%%%%%%%%%%%%%%%%%%%%%%%%%%%%%%%%%%%%%%%%%%%%%%%%%%%%%%%%%%%%%%%%%%%%%%%%%%%%%%%%%%%%%%%%%%%%%%%%%%%%%%%%%%%%%%%%%%%%
%%%%%%%%%%%%%%%%%%%%%%%%%%%%%%%%%%%%%%%%%%%%%%%%%%%%%%%%%%%%%%%%%%%%%%%%%%%%%%%%%%%%%%%%%%%%%%%%%%%%%%%%%%%%%%%%%%%%%%%%%%%%%%%%%%%%%%%%%%%%%%%%%%%%%%%%%%%%%%

\appendix

\section{Spheroidal harmonics and their expansion in spherical harmonics}\label{apdx:spectral-decomp}

The spheroidal harmonics $S_{\ell m}(\theta;\sigma^2)e^{im\phi}$ satisfy the differential equation
\begin{eqnarray}
\left[\frac{1}{\sin\theta}\pdiff{}{\theta}\left(\sin\theta\pdiff{}{\theta}\right) + \left(\lambda_{\ell m} -\sigma^2\cos^2\theta - \frac{1}{\sin^2\theta}\spdiff{}{\phi}\right)\right]S_{\ell m}(\theta;\sigma^2)e^{im\phi} &=& 0 \c \label{eq:spheroidaldiff}
\end{eqnarray}
where the constant parameter $\sigma^2$ is the spheroidicity. The functions $S_{\ell m}(\theta;\sigma^2)e^{im\phi}$ are called  {\it oblate} or {\it prolate} spheroidal harmonics, depending on whether $\sigma^2$ is negative or positive, respectively. A useful and efficient method for calculating the spheroidal harmonics is via decomposition in spherical harmonics. This method is doubly useful in our case, as it automatically generates the spherical-harmonic data required as input for the mode-sum formula. 

The expansion of a given spheroidal harmonic as a series of spherical harmonics, for given $m$, takes the form
\begin{eqnarray}
    S_{\ell m}(\theta;\sigma^2)e^{im\phi} = \sum_{l=l_{\text{min}}}^\infty b_{\ell m}^l(\sigma^2) Y_{l m}(\theta,\phi)        \c
\end{eqnarray}
where $l_{\text{min}} = |m|$. In order to calculate the coefficients $b_{\ell m}^l$ we substitute this expansion into equation (\ref{eq:spheroidaldiff}). Noting that the $Y_{lm}$ satisfy (\ref{eq:spheroidaldiff}) when $\sigma=0$ with $\lambda_{l m} = l(l+1)$, we get
\begin{eqnarray}\label{eq:coefficientSum}
\sum_{l=l_{\text{min}}}^\infty b_{\ell m}^l [ \sigma^2 \cos^2\theta + l(l+l) ] Y_{l m} = \lambda_{\ell m} \sum_{l=l_{\text{min}}}^\infty b_{\ell m}^l Y_{l m}        \p
\end{eqnarray}
Next we multiply the above expression by $Y_{\ell m}^*$ and integrate over the sphere. The resulting inner products are given by  
\begin{eqnarray}
    \oint Y_{\ell m}^* Y_{lm}\, d\Omega &=& \delta_{\ell l}\c       \\
    \oint Y_{\ell m}^* \cos^2\theta\, Y_{lm} d\Omega &=& \frac{1}{3}\delta_{\ell l} + \frac{2}{3}\sqrt{\frac{2l+1}{2\ell+1}}\langle l,2,m,0|\ell,m\rangle\langle l,2,0,0|\ell,0\rangle \equiv k^l_{\ell m}\p          \label{eq:spheroidal-decomp-stage}
\end{eqnarray}
Here the numbers $\langle j_1, j_2, m_1, m_2 | j m\rangle$ are standard Clebsch-Gordan coefficients, the form of which implies that $k^l_{\ell m}\neq 0$ only for $l\in \{\ell-2,\ell-1,\ell,\ell+1,\ell+2\}$.
Consequently, Eq.\ (\ref{eq:coefficientSum}) reduces to the recursion relation
\begin{equation}
   \sigma^2 k^{\ell-2}_{\ell m} b_{\ell m}^{\ell-2}  + \sigma^2 k^{\ell-1}_{\ell m} b_{\ell m}^{\ell-1}  +  [\sigma^2 k^\ell_{\ell m} + l(l+1)]b_{\ell m}^\ell + \sigma^2 k^{\ell+1}_{\ell m} b_{\ell m}^{\ell+1}  + \sigma^2 k^{\ell+2}_{\ell m} b_{\ell m}^{\ell+2}  = \lambda_{\ell m}b_{\ell m}^{\ell}
\end{equation}
for the expansion coefficients $b_{\ell m}^{l}$ (with given $\ell,m$). This can be put in a matrix form, $K{\bf b}=\lambda{\bf b}$ (keeping the indices $\ell,m$ implicit), where $K$ is a known band-diagonal matrix (made up of the known $\sigma^2$ and $k^l_{\ell m}$) and ${\bf b}=(b_{\ell m}^{\ell=1},b_{\ell m}^{\ell=2},\ldots)$. This is a standard eigenvalue problem for the eigenvectors ${\bf b}$ and eigenvalues $\lambda$ (for each $\ell,m$), and the band-diagonality of $K$ makes is readily amenable to numerical treatment. 
This method of obtaining the expansion coefficients $b_{\ell m}^{\ell}$ and spheroidal-harmonic eigenvalues $\lambda_{\ell m}$, which we adopt in this work, follows closely that of Hughes in \cite{Hughes}.

\section{Regularization parameters in Kerr geometry}\label{apdx:regularization-params}

The regularization parameters for the SSF in a generic orbit about a Kerr black hole were calculated by Barack and Ori and given in their Ref.\ \cite{Barack-Ori} (see \cite{Barack-review} for a detailed derivation). For circular, equatorial orbits they reduce to
\begin{eqnarray}
    C_\mu = D_\mu = 0 \c
\end{eqnarray}
and (in Boyer-Lindquist coordinates)
\begin{eqnarray}
    A_r^\pm &=& \mp q^2\Delta^{-1/2}\left(g_{\phi\phi} +\mathcal{L}^2\right)^{-1/2}    \c      \\
    A_t^\pm &=&  A_\theta^\pm  = A_\phi^\pm = 0 \c
\end{eqnarray}
where the metric function $g_{\phi\phi}$ is evaluated on the equatorial orbit. The expression for $B_\mu$ is more complicated. It can be written in the form
\begin{eqnarray}
    B_\mu = q^2 P_{\mu abcd}I^{abcd} \c
\end{eqnarray}
where hereafter Roman indices run over the two Boyer-Lindquist angular coordinates $\theta, \phi$ only. The coefficients $P_{\mu abcd}$ are given by
\begin{eqnarray}
    P_{\mu abcd} = (4\pi)^{-1}[3 P_{\mu d} P_{abc} - (2 P_{\mu ab} + P_{ab\mu})P_{cd}] \c
\end{eqnarray}
where
\begin{eqnarray}
    P_{\alpha\beta}         &\equiv&    g_{\alpha\beta} + u_\alpha u_\beta      \c  \\
    P_{\alpha\beta\gamma}   &\equiv&    (u_\lambda u_\gamma \Gamma^{\lambda}_{\alpha\beta} + g_{\alpha\beta,\gamma}/2)  \c
\end{eqnarray}
with the Kerr connections $\Gamma^\lambda_{\alpha\beta}$ and metric functions $g_{\alpha\beta}$ all evaluated on the equatorial orbit. The quantities $I^{abcd}$ are
\begin{eqnarray}
    I^{abcd} = \int^{2\pi}_0 G(\gamma)^{-5/2}(\sin\gamma)^N(\cos\gamma)^{4-N}\, d\gamma \c
\end{eqnarray}
where 
\begin{eqnarray}
    G(\gamma) \equiv P_{\phi\phi}\sin^2\gamma + 2P_{\theta\phi}\sin\gamma\cos\gamma + P_{\theta\theta}\cos^2\gamma   \c
\end{eqnarray}
and $N \equiv N(abcd)$ is the number of times the index $\phi$ occurs in the combination $(a,b,c,d)$, namely
\begin{eqnarray}
    N = \delta^a_\phi  + \delta^b_\phi + \delta^c_\phi + \delta^d_\phi \p
\end{eqnarray}

The quantities $I^{abcd}$ can be written explicitly in terms of complete elliptic integrals \cite{Barack-Ori,Barack-review}. In the case of a circular, equatorial orbit these expressions become
\begin{eqnarray}\label{Elliptic}
	I^{abcd} = \frac{2(1-w)I^{(N)}_K\hat{K}(w) + I^{(N)}_E \hat{E}(w)}{24P_{\phi\phi}^{5/2}w^4(1-w)^2}	\c
\end{eqnarray}
where $\hat{K}(w) \equiv \int^{\pi/2}_0 (1-w\sin^2 x)^{-1/2}\,dx$ and $\hat{E}(w) \equiv \int^{\pi/2}_0 (1-w\sin^2 x)^{1/2}\,dx$ are complete elliptic integrals of the first and second kind respectively, and
\begin{align}
	w \equiv 1-\frac{P_{\theta\theta}}{P_{\phi\phi}} \p
\end{align}
The coefficients $I^{(N)}_K$ and $I^{(N)}_E$ are given by
\begin{equation}\setlength\arraycolsep{0.1em}
 \begin{array}{rclcl}
I_K^{(0)}	&=&		16 w^2(2-3w)				\c\qquad 	&&I_E^{(0)}	=		64 w^2(2w - 1)				\c		\\
I_K^{(1)} 	&=&		I_E^{(1)} = 0				\c\qquad	&&I_K^{(2)}	=		32 w^2(w - 1)				\c		\\
I_E^{(2)}	&=&		32 w^2(w^2 - 3w + 2)		\c\qquad	&&I_K^{(3)} =		I_E^{(3)} = 0				\c		\\
I_K^{(4)}	&=&		-16 w^2(w^2+w-2)			\c\qquad	&&I_E^{(4)}	=		-64 w^2(w^3 - w^2 - w+1)	\p
 \end{array}
\end{equation}

\section{Boundary conditions for the radial scalar-field equation}\label{apx:bcs-recursion-explicit}

In order to derive recurrence relations for the asymptotic expansion coefficients $c_l^{\infty}$ and $c_k^{eh}$ in Eqs.\ (\ref{eqn:infinity-bc-series}) and (\ref{eqn:horizon-bc-series}), we substitute these equations into the homogeneous part of the radial equation (\ref{eq:rsradialeqn}). By comparing the coefficients of $r^{-k}$ (at infinity) or $(r-r_+)^k$ (at the event horizon) we obtain 5- and 6-term recurrence relations for $c^{\infty}_{k>0}$ and $c^{eh}_{k>0}$, respectively. Setting $c^{\infty,eh}_{k<0}=0$ and  $c^{\infty,eh}_{k=0}=1$ determines all coefficients $c^{\infty,eh}_{k>0}$ in a recursive fashion.

Explicitly, the above recurrence relations are given by
\begin{eqnarray}
    \sum^5_{i=0} f^\infty_i c_{k-i}^\infty = 0, \qquad \sum^6_{i=0}f^{eh}_i c_{k-i}^{eh} = 0        \c
\end{eqnarray}
where the various coefficients $f_i^{\infty}$ and $f_i^{eh}$ read
\begin{widetext}
\begin{eqnarray}
f^\infty_0      &=&     -2 k \omega_m i                                                                                                          \c                          \nonumber\\
f^\infty_1      &=&     k^2-\lambda_{\ell m} + \omega_m(a^2\omega_m - 4iM) + k(4iM\omega_m-1)                                      \c                          \nonumber\\
f^\infty_2      &=&     2[ia^2(2-k)\omega_m + M(a^2\omega_m^2 - 2am\omega_m -2k^2 + 5k - 3 + \lambda_{\ell m} )]      \c                          \nonumber\\
f^\infty_3      &=&     4(k-2)^2 M^2 - a^2(\lambda_{\ell m} - 2k^2 + 8k - 8 - m^2)                           \c                          \nonumber\\
f^\infty_4      &=&     -2a^2M(2k^2 - 11k + 15)                                                                                    \c                          \nonumber\\
f^\infty_5      &=&     a^4 \left(k^2-7 k+12\right)                                                                                             \c                          
\end{eqnarray}
\end{widetext}
\begin{widetext}
\begin{eqnarray}\label{BCin}
f_0^{eh} 	&=& 	a^4 \left(k^2-3 k+2\right)		\nonumber\\
			& &		+a^2 r_+ \left[M \left(-12 k^2+24 k+2 r_+^2 \omega_m^2-6\right)+r_+ \left(12 k^2+k \left(-18-8 i \gamma_m  r_+\right)-\lambda_{\hat{l}m} +m^2+r_+^2 \omega_m^2+2\right)\right]	\nonumber\\
			& &		-4 a m M r_+^3 \omega_m +r_+^2 \left[4 \left(6 k^2-9 k+1\right) M^2+2 M r_+ \left(-20 k^2+2 k \left(12+5 i \gamma_m  r_+\right)+\lambda_{\hat{l}m} -1\right)\right] \nonumber\\ 
			& &		-r_+^4 \left[-15 k^2+3 k \left(5+4 i \gamma_m  r_+\right)+\lambda_{\hat{l}m} +r_+^2 \left(\gamma_m^2-\omega_m^2\right)\right]	\c	\nonumber\\
f_1^{eh} 	&=& 	-2 \left\{a^2 \left[M \left(2 k^2-9 k-3 r_+^2 \omega_m^2+10\right)+r_+ \left((k-1) \left(7+6 i \gamma_m  r_+\right)-4 (k-1)^2+\lambda_{\hat{l}m} -m^2-2 r_+^2 \omega_m^2-2\right)\right]\right. \nonumber\\
			& &		\left.+6 a m M r_+^2 \omega_m +r_+ \left[-2 \left(4 k^2-15 k+13\right) M^2+M r_+ \left((k-1) \left(-26-20 i \gamma_m  r_+\right)+20 (k-1)^2-3 \lambda_{\hat{l}m} +3\right)\right]\right.	\nonumber\\
			& &		\left.+r_+^3 \left[5 (k-1) \left(2+3 i \gamma_m  r_+\right)-10 (k-1)^2+2 \lambda_{\hat{l}m} +3 r_+^2 \left(\gamma_m^2-\omega_m^2\right)\right]\right\}	\c \nonumber\\
f_2^{eh}	&=&		a^2 \left[(k-2) \left(-4-8 i \gamma_m  r_+\right)+2 (k-2)^2-\lambda_{\hat{l}m} +m^2+6 M r_+ \omega_m^2+6 r_+^2 \omega_m^2+2\right]		\nonumber\\
			& & 	-12 a m M r_+ \omega_m +2 M r_+ \left[-10 k^2+k \left(54+20 i \gamma_m  r_+\right)+3 \lambda_{\hat{l}m} -40 i \gamma_m  r_+-71\right]	\nonumber\\
			& &		+4 (k-3)^2 M^2-r_+^2 \left[5 (k-2) \left(3+8 i \gamma_m  r_+\right)-15 (k-2)^2+3 \left(2 \lambda_{\hat{l}m} +5 r_+^2 \left(\gamma_m^2-\omega_m^2\right)\right)\right]	  \c \nonumber\\
f_3^{eh}	&=&		-2 \left\{M \left[-a^2 \omega_m^2+2 a m \omega_m +(k-3) \left(-3-10 i \gamma_m  r_+\right)+2 (k-3)^2-\lambda_{\hat{l}m} +1\right]\right.	\nonumber\\
			& &		\left.+(k-3) \left(i a^2 \gamma_m +15 i \gamma_m  r_+^2+3 r_+\right)+2 r_+ \left[-a^2 \omega_m^2+\lambda_{\hat{l}m} +5 r_+^2 \left(\gamma_m^2-\omega_m^2\right)\right]-3 (k-3)^2 r_+\right\}	\c \nonumber\\
f_4^{eh}	&=&		a^2 \omega_m^2+i (k-4) \left(4 \gamma_m  M-12 \gamma_m  r_++i\right)+(k-4)^2-\lambda_{\hat{l}m} -15 r_+^2 \left(\gamma_m^2-\omega_m^2\right)	\c \nonumber\\
f_5^{eh}	&=&		-2 i \gamma_m  (k-5)-6 \gamma_m^2 r_++6 r_+ \omega_m^2 \c \nonumber \\
f_6^{eh}	&=&		\omega_m^2-\gamma_m^2 \p
\end{eqnarray}
\end{widetext}
Note, an earlier version of this work contained an incorrect recursion relation for the inner boundary conditions. This did not effect the numerical results, as we placed our inner boundary sufficiently close to the horizon that only the (correct) leading term in the expansion (\ref{eqn:horizon-bc-series}) contributed.

\bibliographystyle{apsrev}
\bibliography{Warburton}

\end{document}